\documentclass[12pt,letterpaper]{report}
\usepackage{natbib}
\usepackage{geometry}
\usepackage{fancyhdr}
\usepackage{afterpage}
\usepackage{graphicx}
\usepackage{amsmath,amssymb,amsbsy}
\usepackage{dcolumn,array}
\usepackage{tocloft}
\usepackage{asudis}
\usepackage{csquotes}
\usepackage{algorithm}
\usepackage{algorithmic}
\usepackage[toc,page]{appendix}
\usepackage[table,xcdraw]{xcolor}
\usepackage[english]{babel}

\begin{document}
\pagenumbering{roman}
\degreeName{Master of Science}
\paperType{Thesis}
\title{Multi-Perspective Semantic Information Retrieval in the Biomedical Domain}
\author{Samarth Rawal}
\degreeName{Master of Science}
\defensemonth{April}
\gradmonth{May}
\gradyear{2020}
\chair{Chitta Baral}
\memberOne{Murthy Devarakonda}
\memberTwo{Saadat Anwar}

\maketitle
\doublespace
\begin{abstract}

Information Retrieval (IR) is the task of obtaining pieces of data (such as documents  or snippets of text) that are relevant to a particular query or need from a large repository  of information. IR is a valuable component of several downstream Natural Language Processing (NLP) tasks, such as Question Answering. Practically, IR is at the heart of many widely-used technologies like search engines.

While probabilistic ranking functions, such as the Okapi BM25 function, have been utilized in IR systems since the 1970's, modern neural approaches pose certain advantages compared to their classical counterparts. In particular, the release of BERT (Bidirectional Encoder Representations from Transformers) has had a significant impact in the NLP community by demonstrating how the use of a Masked Language Model (MLM) trained on a considerable corpus of data can improve a variety of downstream NLP tasks, including sentence classification and passage re-ranking. 

IR Systems are also important in the biomedical and clinical domains. Given the continuously-increasing amount of scientific literature across biomedical domain, the ability find answers to specific clinical queries from a repository of millions of articles is a matter of practical value to medics, doctors, and other medical professionals. Moreover, there are domain-specific challenges present in the biomedical domain, including handling clinical jargon and evaluating the similarity or relatedness of various medical symptoms when determining the relevance between a query and a sentence.

This work presents contributions to several aspects of the Biomedical Semantic Information Retrieval domain. First, it introduces Multi-Perspective Sentence Relevance, a novel methodology of utilizing BERT-based models for contextual IR. The system is evaluated using the BioASQ Biomedical IR Challenge. Finally, practical contributions in the form of a live IR system for medics and a proposed challenge on the Living Systematic Review clinical task are provided.
\end{abstract}
\dedicationpage{I would like to dedicate this work to my family, friends, and mentors who have always supported me.}
\begin{acknowledgements}
I would like to thank Dr. Chitta Baral for his support and guidance over the past six years, both as a research advisor and mentor. His steady encouragement has been invaluable in enabling me to pursue research and participate in projects related to my interest in clinical Natural Language Processing.

Next, I would like to thank Dr. Saadat Anwar for his mentorship and advice over the past several years. From my early years as an undergraduate student up until now, he has offered valuable feedback and guidance that has helped me grow.

I would also like to thank Dr. Murthy Devarakonda for serving as a mentor to me and introducing me to many ways computer science can have clinical applications.

Finally, I would like to thank my family and friends, including my parents and brother, as well as my colleagues at Arizona State University. Specifically, I would like to thank Arpit Sharma, who was one of my earliest student mentors and taught me much of what I know today about conducting research. I would also like to thank my friends Ishan Shrivastava, Aurgho Bhattacharjee, Arindam Mitra, Sanjay Narayana, Swaroop Mishra, Pratyay Banerjee, Kuntal Pal, Shailaja Sampat, and many others for their help and encouragement throughout my time at the lab.

\end{acknowledgements}
\tableofcontents
\addtocontents{toc}{~\hfill Page\par}
\newpage
\addcontentsline{toc}{part}{LIST OF TABLES}
\renewcommand{\cftlabel}{Table}
\listoftables
\addtocontents{lot}{Table~\hfill Page \par}
\newpage
\addcontentsline{toc}{part}{LIST OF FIGURES}
\addtocontents{toc}{CHAPTER \par}
\renewcommand{\cftlabel}{Figure}
\listoffigures
\addtocontents{lof}{Figure~\hfill Page \par}
\doublespace
\pagenumbering{arabic}
\chapter{INTRODUCTION AND MOTIVATION}

Perhaps most widely utilized in the form of search engines, the ability to take a query and use it to find relevant information from a repository of knowledge is essential to effectively making use of large amounts of data. The field of study that deals with finding and ranking relevant information based on certain criteria is Information Retrieval (IR), which utilizes many components from the field of Natural Language Processing (NLP), particularly when dealing with textual data such as webpages or research papers. NLP can be considered to be a subfield of Artificial Intelligence (AI) that deals with the semantic interpretation of natural language by machines. Over the decades, several NLP statistical metric- and machine learning-based techniques, such as term frequency–inverse document frequency, bag-of-words, and deep neural networks, have been applied to IR systems with varying degrees of success.

Information Retrieval systems have considerable research and practical value. IR systems are used in NLP tasks like question answering, document classification, and automatic document summarization. From a practical standpoint, systems that can accurately locate relevant data can have many domain-specific uses. For instance, there is considerable real-world value in being able to accurately and automatically locate relevant scientific literature from a repository of millions of papers that can answer a particular clinical query. An effective IR system can offset considerable manual human review of data and allow users to focus on information that is relevant. From casual web searching to specific tasks like clinical study results lookup, there is significant interest and value in the development of accurate and reliable IR systems.

However, discussing such a system begs the question, \emph{what does it mean for a sentence or document to be relevant to a query?} After all, a piece of information's "relevance" to a query is rarely a simple binary value. For instance, two sentences may both be relevant to a query, but one could be \emph{more relevant} than the other by containing more complete information. Part of the objective of IR is to determine what constitutes "relevance" and how to locate information that matches this. 

In this work, we present a semantic Information Retrieval system for the biomedical domain which incorporates elements from several NLP tasks such as Sentence Relevance, Semantic Textual Similarity, and Semantic Information Availability. We demonstrate the effectiveness of this system in retrieving and ranking relevant documents and sentences from the MEDLINE/PubMed Baseline through the BioASQ Challenge. Additionally, we present multiple applications of the work described above, in the form of (1) an interactive and functional interface, and (2) a proposal of multiple challenges with respect to application of such semantic IR systems to the Living Systematic Review workflow, a clinical workflow which is currently done entirely manually in practice.

\section{Information Retrieval}

Information Retrieval (IR) is an active area of research with significant downstream practical and research-related applications, especially when paired with Question Answering (QA). The goal of an IR system is to take in a query and locate from a data repository pieces of information that are most relevant to the query. Such a system therefore requires the ability to parse and interpret a query (employing methods from the subfield of QA), as well as to evaluate and rank the relevance of data in the repository with respect to the query. This work will specifically focus on IR in the textual domain.

The meaning and nature of a "relevant" document given a query can depend on the domain and the task for which IR system is being used. For instance, in the biomedical domain, if a medical professional submits a query about the efficacy of a specific drug on a particular illness, they would likely be looking for the results of clinical trials, rather than general information about a particular drug (which could be interpreted as being relevant to the query, but would not be relevant with respect to the likely intent behind the search). 

In order to create an IR system that searches over a particular data repository (such as a repository of text files), the data in the repository needs to be converted into a representation that facilitates efficient searching and ranking. Once such an index is constructed, a ranking algorithm is used to locate the most relevant documents in the index given a particular query. A brief overview of the various indexing methods and ranking algorithms is presented in the subsequent Chapter. 

\section{BioASQ Challenge}
The BioASQ Challenge is a semantic indexing and question answering competition for the biomedical domain \citep{bioasqoverview}. In this work, we will be focusing on Task B, Phase A of this challenge, specifically on the document retrieval and sentence retrieval components of this challenge. The BioASQ dataset consists of a series of queries. For each query, up to 10 gold "relevant" documents with respect to the query are provided in the form of their PubMed URLs, and up to 10 "relevant" snippets of text with respect to the query (most commonly single sentences), which have been obtained from the gold documents are provided. A "document" consists of the title and abstract of a paper on PubMed. The dataset from which these documents are obtained is the MEDLINE/PubMed Baseline repository, which consists of roughly 28 million articles (where an article consists of a paper's document and abstract).

For this particular challenge, teams receive a list of queries. For each query, the system must first obtain up to 10 most relevant documents, and from those documents, up to 10 most relevant snippets of text. Thus, the snippet retrieval task can be considered a "downstream" task of the document retreival task. The results are evaluated on a variety of metrics that include the F1 score and Mean Average Precision (MAP) of the document and sentences. 

Table \ref{relevant-nonrelevant-samples} provides examples of relevant and nonrelevant sentences, taken from documents in the MEDLINE/PubMed 2018 Baseline dataset, given a particular query. The samples in the table were obtained from the BioASQ Challenge, and "relevant" samples are defined as the Top 10 most relevant sentences from a particular query per the BioASQ Gold Dataset; the "nonrelevant" sentences are those outside the Top 10. 

\begin{table}[]
\begin{tabular}{|l|l|}
\hline
\textbf{Query} & What is the effect of TRH on myocardial contractility? \\ \hline
\textbf{\begin{tabular}[c]{@{}l@{}}Relevant\\ Samples\end{tabular}} & \begin{tabular}[c]{@{}l@{}}• Acute intravenous administration of TRH to rats with \\ ischemic cardiomyopathy caused a significant increase in heart \\ rate \\• TRH can enhance cardiomyocyte contractility in vivo\\ • TRH in the range of 0.1-10 mumol/l was found to exert \\a positive  inotropic effect on cardiac contractility.\\ • Thyrotropin-releasing hormone (TRH) improved mean arterial \\ pressure (MAP) and myocardial contractility (dp/dtmax, \\ -dp/dtmax, Vpm, and Vmax).\\ • TRH improves cardiac contractility, cardiac output,\\ and hemodynamics.\end{tabular} \\ \hline
\textbf{\begin{tabular}[c]{@{}l@{}}Nonrelevant\\ Samples\end{tabular}} & \begin{tabular}[c]{@{}l@{}}• These data suggest that 5-HT is an important transcriptional \\ regulator  of the cardiac TRH gene.\\ • The effects of thyrotropin-releasing hormone (TRH) and the \\TRH-analogs, 4-fluoro-Im-TRH (4-F-TRH) and, \\ 2-trifluoromethyl-Im-TRH (2-TFM-TRH), on the cardiovascular \\ system and prolactin (PRL) release were examined in conscious \\ rats.\\ • Thyrotropin-releasing hormone (TRH) has been shown to \\ be scattered throughout the gastrointestinal tract.\\ • It is concluded that the enhancement by TRH of \\ indomethacin-induced  gastric lesions is due to a combination \\ of the central and peripheral actions of the ulcerogenic agents.\end{tabular} \\ \hline
\end{tabular}
\caption[Relevant and Nonrelevant Samples, Given a Query]{Relevant and nonrelevant samples, given a query.}
\label{relevant-nonrelevant-samples}
\end{table}

\section{Research Value}
Having the ability to effectively access biomedical information has importance of practical value in the clinical field. For instance, clinical sites in developing nations, offshore sites, or in areas with limited resources, may not have the resources to maintain an up-to-date repository of medical best-practices for a variety of situations to easily access when needed. In areas such as this, having an Information Retrieval and Question Answering system that can accept a clinical or biomedical query and automatically obtain sentences or documents that can address such a query. Doing so would obviate the need for extensive manual review of hundreds of new scientific literature, and allow professionals to concentrate on only the most relevant literature and spend greater time on making clinical decisions.

\section{Research Evaluation}
The approach is evaluated on the BioASQ 6B-Phase A challenge , specifically on the document and sentence retrieval tasks. The official evaluation script is used, which expects a predictions file in the same format as the gold standard file (JSON files) and outputs a series of metrics, including Precision, Recall, F1, and Mean Average Precision for both document and sentence retrieval tasks. The official evaluation metric for this particular challenge is the Mean Average Precision (MAP) metric. The evaluation script will be used to obtain all of the described metrics.

\section{Contributions}
This work presents a series of contributions related to semantic Information Retrieval in the biomedical domain:

\begin{enumerate}
    \item \textbf{Contextual Multi-Perspective Sentence Relevance} 
    
    The main contribution of this work is a novel approach for conducting Sentence Ranking and Document Ranking for application in tasks such as Information Retrieval. This approach involves the evaluation of multiple "perspectives" related to query-sentence pair relevance and the fusion of these perspectives to form a more rounded ranking score. The development of this methodology, as well as successes and challenges encountered in various iterations of the process, are presented.
    
    \item \textbf {BioASQ Document and Sentence Ranking Challenges} 
    
    The Multi-Perspective Sentence Relevance system was evaluated on the BioASQ 6B and 7B Phase A tasks. The results and accompanying analysis are presented.
    
    \item \textbf{Interactive Semantic IR Demo for Medics} 
    
    An interactive demo, representing a valuable application of Semantic Information Retrieval in retrieving information from medical handbooks, has been developed as part of this work. Details about the demo and corresponding Application Programming Interface (API) are provided.
    
    \item \textbf{Semantic IR Applied to Living Systematic Review} 
    
    Background information about clinical Systematic Reviews, as well as multiple challenges regarding the application of Semantic Information Retrieval to Living Systematic Reviews, is presented.
    
\end{enumerate}

\section{Structure of Thesis}
This work will first review some background and existing work on the topic of Information Retrieval and Semantic Information Retrieval. Next, it will cover the novel Semantic IR approach taken in this work, including the intuitions behind it, formulations, and training process. Next, the system's performance in the BioASQ Document and Sentence Ranking challenges will be reported and analyzed. A practical application of the system in the form of an IR application for medics will then be examined. Next, a new challenge will be proposed around the Living Systematic Review clinical task. Finally, some concluding remarks and areas of future work will be detailed.
\chapter{BACKGROUND AND RELATED WORK}

In this Chapter, we will discuss various techniques utilized for Information Retrieval, ranging from ranking models developed several decades ago to modern, deep learning-based techniques. In Section 2.1 we will briefly review IR methodologies developed and used over the past several decades. In Section 2.2, we will discuss recent neural breakthroughs in NLP and their applicability to IR. In Section 2.3, we will discuss recent semantic IR systems that leverage neural methodologies for document and sentence retrieval tasks. In Section 2.4, we will review the application of the previously-mentioned techniques to specifically the biomedical domain.

\section{Traditional Information Retrieval Methodologies}
Variants of automated information retrieval systems have been implemented for the past several decades. At the heart of being able to retrieve documents is creating effective representations of them. Historically, models such as the Vector Space model and Probabilistic model, which rely on factors including term frequency, inverse document frequency, document length, have been used to come up with document representations \citep{mitra2000information}.

Once the documents have been converted from text into a numerical or vector-based representation, a \emph{ranking function} needs to sort the documents in terms of relevance to a particular query. The Okapi BM25 ranking function, a derivative of the Probabilistic retrieval model, is one of the most well-known and widely-used ranking functions in IR, and is a bag-of-words retrieval function that also relies on term frequency and inverse document frequency (TF-IDF).

\section{Neural Breakthroughs in NLP}

BERT (Bidirectional Encoder Representations from Transformers) is a language representation model that, along with similar types of models, has empirically been shown to yield state-of-the-art results in several NLP tasks, such as on the GLUE benchmark, and on tasks related to IR like Question Answering \citep{devlin2018bert}. BERT and related models are able to achieve significant performance gains in part due to the large corpora of data they are trained on. Moreover, the Transformer neural network architecture \citep{vaswani2017attention}, which BERT is built on, facilitates deep relationships between individual tokens in an input, thereby allowing the model to gain better contextual awareness of the relationships present in the input.

\section{Recent Semantic IR Systems}
Apache Lucene is an open-source search library which is widely used for developing and deploying IR systems, both in research and commercial applications \citep{lucene2010apache}. The software is able to turn raw data repositories into indices containing the data represented in a form which allows for efficient searching and ranking. Moreover, common ranking and searching algorithms, such as the aforementioned Okapi BM25 algorithm, are implemented in the software. Anserini is an IR toolkit built on top of Lucene that facilitates easier experimentation and reproduction of results when using bag-of-words ranking models like BM25 \citep{yang2018anserini}. Anserini also has a Python interface named Pyserini.

Recently, several IR systems have been developed that take advantage of the ability of BERT to capture contextual information. Because running BERT models is computationally expensive compared to ranking algorithms like BM25, these systems generally use BM25 or similar algorithms to perform a "coarse" level retrieval to narrow down candidates from the order of millions of documents to tens or hundreds, and then carry out finer reranking via BERT.

One such system is Birch, which utilizes a combination of Anserini for coarse ranking and BERT for fine ranking \citep{yilmaz2019birch}. BERT performs top Document Retrieval via Anserini and further re-ranking of the document list via BERT.

One key insight made in \cite{nogueira2019passage}, which will be explored further in the context of the work detailed in this paper, is the notion that the Semantic Rank of a particular document can be represented by the weighted sum of the document's 3 most relevant Sentences. In this way, a basic formulation of Document Ranking via Sentence Ranking can be made. This idea will be discussed further in subsequent Chapters.

\section{Semantic IR in the Biomedical Domain}
There have been many innovations specifically in the biomedical and clinical NLP domains. Some widely-used pretrained and finetuned variants of BERT are BioBERT \citep{10.1093/bioinformatics/btz682biobert} and NCBI BlueBERT \citep{peng2019transferbluebert}. Various participants in BioASQ Challenges have utilized variants of BERT, such as BioBERT or BERT-Large, for the Sentence Ranking components.

Other participants in the BioASQ Challenge, have also utilized BERT-based systems for document and/or snippet retrieval purposes. Traditionally, such IR systems have consisted of two independent modules, a document ranking and sentence ranking module, with the overall system operating as a "pipeline". This formulation matches the BioASQ Challenge, where Sentence Retrieval is a downstream task of the Document Retrieval challenge. However, some participants, such as the team from Athens University of Economics and Business, have developed methodologies to jointly rank and retrieve documents and sentences, to successful results \citep{pappas2019aueb}. As utilizing BERT and BERT-like neural models can be computationally expensive, efficient ranking algorithms like the previously-mentioned BM25 algorithm are generally used to retrieve a smaller subset of potentially relevant documents from the full repository of about 25 million documents. The more contextual, computationally-expensive algorithms are then run on the smaller subset of documents rather than on the full repository. 
\chapter{COMPONENTS OF SEMANTIC BIOMEDICAL INFORMATION RETRIEVAL}

Having briefly covered the foundations of Information Retrieval systems, along with relevant prior work, we will discuss our IR system designed for Document and Sentence Retrieval in the biomedical domain. We have seen from previous literature that a combination of traditional and neural network-based approaches can incorporate greater contextual awareness in IR algorithms, leading to better results. In this section, we will describe our semantic Information Retrieval system, including the various iterations (both successful and unsuccessful) of the algorithm throughout the course of its development.

The main motivating objectives behind the development of this system were: 
\begin{itemize}
    \item The BioASQ Information Retrieval + Question Answering Challenge
    \item An Interactive Semantic IR System for Medics System
    \item Performing automated Living Systematic Reviews
\end{itemize}

This Chapter will detail the development of the system and its fundamental components. We will leave evaluation and task-specific descriptions of system to subsequent Chapters.

\section{General System Architecture}
The primary focus of research and development was on a Sentence Ranking module, as that module was directly applicable to all three of the main motivating objectives. Because the BioASQ Challenge requires Document Retrieval and Ranking to be performed upstream of Sentence Ranking, a Document Ranking module -- formulated using the Sentence Ranking module -- was also developed. Because the Sentence Ranking module is central to all three objectives, including in the Document Retrieval task, we will first discuss that module.

\begin{figure}[h]
\includegraphics[width=15cm]{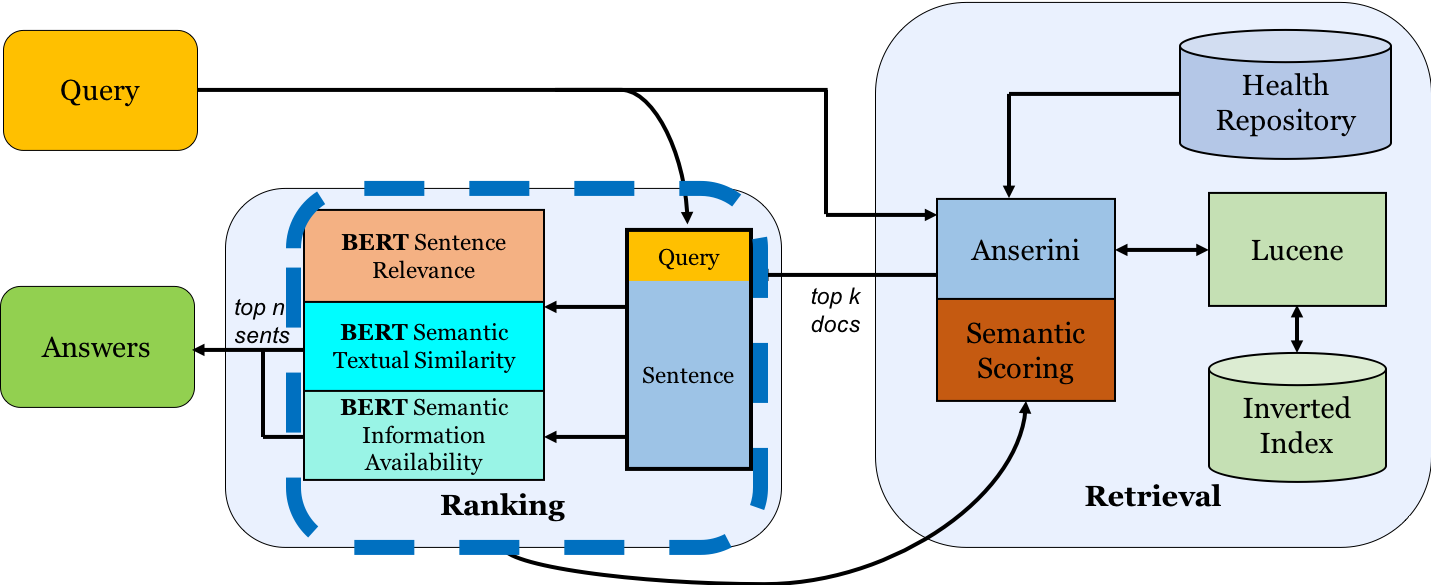}
\caption[Overall Semantic IR System Architecture]{Overall Semantic IR system architecture}
\label{fig:overallarchitecture-1}
\centering
\end{figure}

Figure \ref{fig:overallarchitecture-1} provides a high-level overview of the system as a whole as part of the BioASQ Challenge, with the Sentence Ranking module emphasized. The diagram illustrates the use of the Sentence Ranking module in both the Document Retrieval and Sentence Ranking tasks. The first task of the system for the BioASQ Challenge is to take a query and retrieve the top $k$ relevant documents out of roughly 29 million candidates in the MEDLINE/Pubmed Baseline, where $k$ is an integer that is set by the user depending on the objectives (for the BioASQ Challenge, $k$ is up to 10). Given the top $k$ relevant documents along with the original query, the second part of the system will break the documents down into individual sentences and perform semantic ranking on the sentences to identify the top $n$ most relevant sentences given the query, where $n$ is again an integer that is set by the user (for the BioASQ Challenge, $n$ is up to 10).

This Chapter will discuss the development of the Sentence Ranking module first, followed by the Document Retrieval component. 

\section{Sentence Ranking}
The Sentence Ranking component of the IR system is comprised of multiple sub-systems. Each of these systems was developed and refined separately, and then further refined as part of the Sentence Ranking component. Thus, the following section will cover the various sub-systems of the Sentence Ranking module, along with challenges in the development of them, and then discuss their fusion.

We randomly divide the BioASQ training data into a training and development set, of 1742 and 433 queries, respectively. Each query contains up to 10 ranked documents and up to 10 ranked snippets (which for the most part are single sentences), which we will refer to as the "training set" and "development set" in the subsequent subsections.

\subsection{Sentence Relevance}
\begin{figure}[t]
\includegraphics[width=15cm]{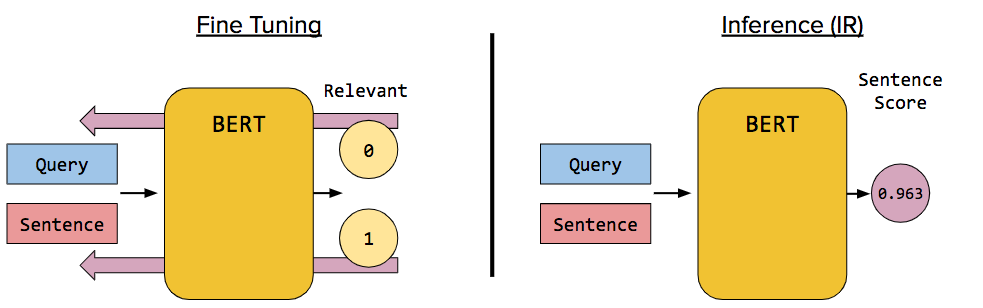}
\caption[Overall Sentence Relevance Model Fine-Tuning and Inference-Time Methodologies]{Overall Sentence Relevance model fine-tuning and inference-time methodologies.}
\label{fig:sentrel_method}
\centering
\end{figure}

As noted in previous literature, utilizing BERT-based architectures for Sentence Relevance can provide comparatively effective results for sentence relevance ranking. This work follows the approach detailed by Nogueira et al. in repurposing BERT as a Passage Reranker \citep{nogueira2019passage}. A BERT-Large model (with 24 layers, 16 attention heads, and 340 million parameters) is trained on a binary classification task of classifying a query-sentence pair as a "relevant" pair or "not relevant" pair. During inference, the prediction probability is taken as the Sentence Ranking score; sentences with higher scores are considered to be more relevant. Figure \ref{fig:sentrel_method} provides a high-level overview of this process for both fine-tuning and inference in IR systems. As noted by the authors in \cite{nogueira2019passage}, this model was surprisingly effective across various domains; specifically, they demonstrated its effectiveness across general English training and Twitter domain IR performance \citep{nogueira2019passage}. 

We start with a BERT model that has been fine-tuned on a binary classification task of whether a particular sequence was relevant to a given query or not, using the Microsoft MS MARCO dataset \citep{nguyen2016msmarco}. The model weights for such a finetuned version of this model, detailed in the paper, have been made available by the authors on GitHub \citep{nogueira2019passage}. The authors have trained the model on a dataset of 12.8 million \texttt{(query, sentence)} pairs from the MS MARCO dataset.

Using the BioASQ training data, we create a biomedical binary sentence relevance dataset in the same format in order to further finetune the MS MARCO-finetuned model using data from the biomedical domain. We generate binary training samples using the training set and evaluate using binary samples generated from the development set. The training and development binary sentence relevance datasets are 221K and 55K samples, respectively. We further finetune the model on this domain-specific binary sentence relevance task, evaluating on the development set. Examples from this dataset are shown in Table \ref{tab:sentence-relevance-examples}.

We find that, as the model has already been finetuned once on the MS MARCO dataset, only a small amount of further finetuning -- under 1 epoch -- is necessary to achieve a peak in the model's performance on both the binary sentence relevance score and the BioASQ Sentence Ranking score. Each checkpoint of the model is evaluated on both these measures, and it is found that further finetuning for 1000 checkpoints on the biomedical sentence relevance dataset leads to a small gain in the model's performance compared to the MS MARCO "baseline" model. Moreover, it can be seen that there is a correlation between performance on the binary sentence relevance task -- evaluated using the F1 metric -- and the sentence ranking task -- evaluated using the Mean Average Precision, per the BioASQ Guidelines, in that a higher score on the binary sentence relevance task largely correlates with a higher MAP score in the BioASQ sentence ranking task (when the documents are fixed with the gold input). The hyperparameters used to finetune this model are described in Table \ref{tab:sentrel-hyperparameters}.

\begin{table}[]
\begin{tabular}{|l|}
\hline
\begin{tabular}[c]{@{}l@{}}( Is nintedanib effective for Idiopathic Pulmonary Fibrosis?, \\   In this review, we present the positive results of recently published clinical \\ trials    regarding therapy for IPF, with emphasis on pirfenidone and nintedanib., \\    1  )\end{tabular} \\ \hline
\begin{tabular}[c]{@{}l@{}}( Is nintedanib effective for Idiopathic Pulmonary Fibrosis?,\\   Results will be reported in the first half of 2014.,\\    0  )\end{tabular} \\ \hline
\begin{tabular}[c]{@{}l@{}}(Mutation of which gene is implicated in the familial isolated pituitary \\ adenoma?,   Germline mutations in the aryl-hydrocarbon interacting protein gene \\ are identified     in around 25\% of familial isolated pituitary adenoma kindreds.,\\   1  )\end{tabular} \\ \hline
\begin{tabular}[c]{@{}l@{}}(Mutation of which gene is implicated in the familial isolated \\ pituitary adenoma?,   However, the exact molecular mechanism by which \\ its disfunction promotes tumorigenesis of pituitary is unclear.,\\    0  )\end{tabular} \\ \hline
\end{tabular}
\caption[Examples from the Biomedical Science Relevance Dataset]{Examples from the biomedical Sentence Relevance dataset.}
\label{tab:sentence-relevance-examples}
\end{table}

The data is formulated in the following manner to pass into the BERT model:

\begin{center}
\texttt{[CLS] Query [SEP] Sentence}
\end{center}

where $Query$ and $Sentence$ are the tokenized query and the sentence respectively, $[CLS]$ is the classification token, and $[SEP]$ is the separator token that separates the query and sentence. For example,

\begin{center}
\texttt{[CLS]What is Dravet syndrome?[SEP]Dravet syndrome is a severe form of epilepsy.}
\end{center}

(where the sentences would be tokenized and represented as vectors of numbers rather than alphabetically).

\begin{table}[]
\centering
\begin{tabular}{|l|l|}
\hline
\textbf{learning rate} & 3e-06 \\ \hline
\textbf{seq\_len} & 128 \\ \hline
\textbf{weight\_decay} & 0.01 \\ \hline
\textbf{adam\_epsilon} & 1e-08 \\ \hline
\end{tabular}
\caption[Hyperparamters Used to Finetune Sentence Relevance BERT Model.]{Hyperparameters used to finetune Sentence Relevance BERT model.}
\label{tab:sentrel-hyperparameters}
\end{table}

Like that of MS MARCO task, the loss function of this additional finetuning is a Cross-Entropy loss. In other words, the goal of this process is to try to finetune a BERT model to accurately predict whether or not a particular sentence, given a query, is relevant or not.

During inference time as part of the Sentence Retrieval pipeline, each relevant document retrieved from the Document Retrieval pipeline is split into individual sentences. Each sentence is then tokenized in the same format as described above, and passed through the Sentence Relevance BERT model. Because the BERT model has been trained to classify between two classes (0/not relevant and 1/relevant), the model outputs two values; when passed through a softmax layer, they sum to 1.0 and represent the prediction probability for each of the possible classes. To determine a particular sentence's Sentence Relevance rank, the probability of the 1 class is taken. Thus, each sentence from all the relevant documents will receive a score between 0.0 and 1.0; the sentences will then be sorted by this score. Figure \ref{fig:sentrel_method} provides a high-level overview of the fine-tuning and inference-time processes.

To evaluate the performance of the models on the binary Sentence Relevance dataset, the Precision, Recall, and F1 metrics were used. Scores are shown in Table \ref{tab:binary-sentence-relevance-scores}. 
It was found that the finetuned model at checkpoint of step 1000 was the best performing on this sentence relevance task compared to the original MS MARCO finetuned model and the model finetuned on more data.

\begin{table}[]
\begin{tabular}{|l|c|c|c|}
\hline
\textbf{Model Description} & \textbf{Precision} & \textbf{Recall} & \textbf{F1 Score} \\ \hline
MS MARCO & 0.4154 & 0.6118 & 0.4948 \\ \hline
\begin{tabular}[c]{@{}l@{}}MS MARCO + Bio Sentence Relevance\\ (checkpoint 1000)\end{tabular} & 0.6813 & 0.5466 & 0.6066 \\ \hline
\begin{tabular}[c]{@{}l@{}}MS MARCO + Bio Sentence Relevance\\ (checkpoint 2000)\end{tabular} & 0.7484 & 0.4860 & 0.5893 \\ \hline
\begin{tabular}[c]{@{}l@{}}MS MARCO + Bio Sentence Relevance\\ (checkpoint 3000)\end{tabular} & 0.7580 & 0.4806 & 0.5882 \\ \hline
\end{tabular}
\caption[Model Performances on Biomedical Binary Sentence Relevance Task. Scores for the "1" Class.]{Model performances on biomedical binary sentence relevance task. Scores for the "1" class.}
\label{tab:binary-sentence-relevance-scores}
\end{table}

 We found these results on the binary Sentence Relevance dataset also correlated with performance on the Sentence Ranking task from the BioASQ Challenge. To evaluate on the BioASQ Sentence Ranking task, we fixed the Document Retrieval portion by feeding the system the gold labeled documents, and in this way evaluated the Sentence Ranking by itself. We found that in this task, the clinical Sentence Relevance finetuned at checkpoints 1000 and 3000 performed the highest in terms of the MAP measure. The scores for this are shown in Table \ref{tab:sentence-ranking-bioasq-scores}. 
 
\begin{table}[]
\begin{tabular}{|l|c|c|c|c|c|}
\hline
\textbf{Model Name} & \textbf{\begin{tabular}[c]{@{}c@{}}MPrec\\ snippets\end{tabular}} & \textbf{\begin{tabular}[c]{@{}c@{}}MRec\\ snippets\end{tabular}} & \textbf{\begin{tabular}[c]{@{}c@{}}MF1\\ snippets\end{tabular}} & \textbf{\begin{tabular}[c]{@{}l@{}}MAP\\ snippets\end{tabular}} & \textbf{\begin{tabular}[c]{@{}l@{}}GMAP\\ snippets\end{tabular}} \\ \hline
\begin{tabular}[c]{@{}l@{}}MS MARCO + \\Bio Sent Relevance \\(checkpoint 1000)\end{tabular} & 0.4571 & 0.4630 & 0.4037 & 0.5016 & 0.1173 \\ \hline
\begin{tabular}[c]{@{}l@{}}MS MARCO + \\Bio Sent Relevance \\(checkpoint 2000)\end{tabular} & 0.4576 & 0.4681 & 0.4068 & 0.5000 & 0.1139 \\ \hline
\begin{tabular}[c]{@{}l@{}}MS MARCO + \\Bio Sent Relevance \\(checkpoint 3000)\end{tabular} & 0.4602 & 0.4702 & 0.4087 & 0.5022 & 0.1135 \\ \hline
\end{tabular}
\caption[Model Performances on BioASQ Sentence Ranking Challenge. Evaluated on the Development Dataset Using Gold Relevant Document List.]{Model performances on BioASQ Sentence Ranking challenge. Evaluated on the development dataset using Gold Relevant Document list.}
\label{tab:sentence-ranking-bioasq-scores}
\end{table}

 In short, we found that employing methods similar to the existing work discussed in Chapter 2 provided decent results on the BioASQ Challenge. However, there are a few key issues that remain unaddressed with this method. The central problem with solely relying on a binary Sentence Relevance-based system for Sentence Ranking tasks is that \emph{binary Sentence Relevance is not equivalent to Sentence Ranking by importance}.
 
 There are several aspects of binary Sentence Relevance that make it advantageous for training such systems on. Primarily, there is a significant amount of labeled data for this task, both in general English and in biomedical domains. Moreover, it is relatively simple to generate additional data, both automatically and semi-automatically, by modifying existing datasets, as we did with the BioASQ dataset. However, intuitively speaking, just because a sentence is "relevant" does not mean it is a top-ranked sentence in terms of what a user submitting a query is looking for. In other words, "relevance" is generally not a binary attribute -- there are different degrees of relevance -- two sentences can both be relevant, but one can be more relevant than the other -- that cannot be adequately captured by such a binary formulation.
 
 Additionally, in the BioASQ challenge, sentences and documents are ranked by \emph{confidence of relevance}, rather than just relevance. A different way of interpreting this task is to rank sentences by the amount of "evidence" or "justification" that can be compiled about its relevance. This intuitive understanding leads us to the consideration of other metrics and methodologies that can be used to gather additional "justification" for the relevance of a sentence which can ideally fill the gaps present through the sole utilization of the Sentence Relevance-based methodology.
 
\begin{figure}[t]
\includegraphics[width=15cm]{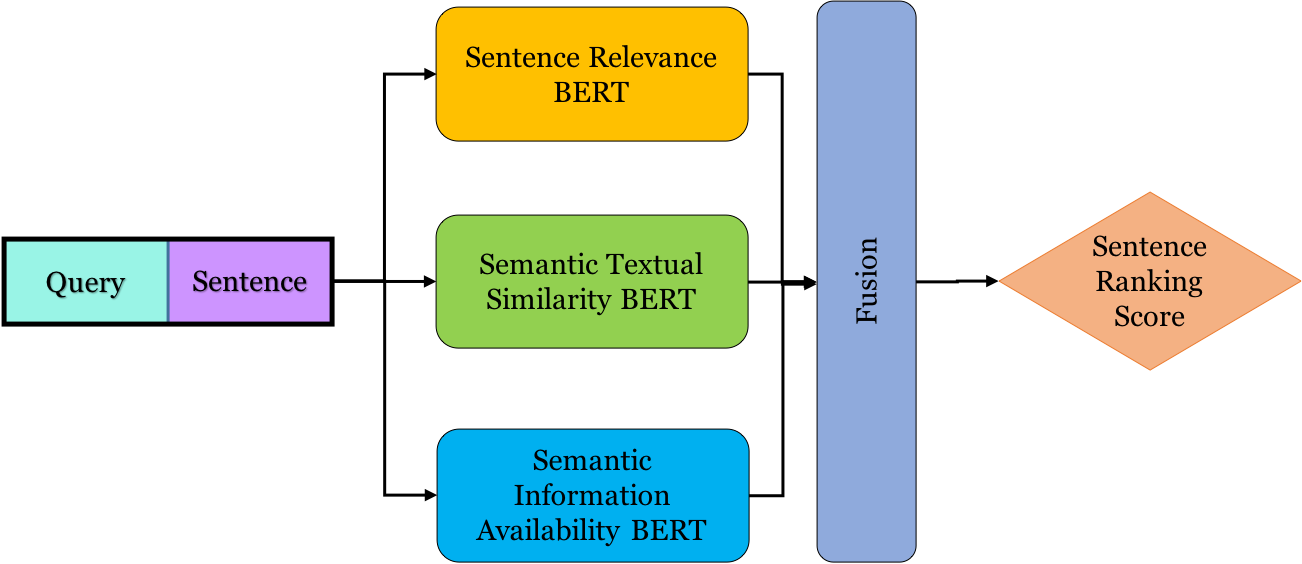}
\caption[High-Level Multi-Perspective Architecture Concept.]{High-level Multi-Perspective architecture concept.}
\label{fig:multiperspective-1}
\centering
\end{figure}
 
 Thus, we propose the fusion of three models trained on different NLP tasks -- Sentence Relevance, Semantic Textual Similarity, and Semantic Information Availability -- to create a joint representation of the relevance of a sentence. We will first describe the remaining two tasks of Semantic Textual Similarity and Semantic Information Availability, then discuss the process of fusing these three representations into a single one. In the following Chapters, we will discuss the empirical results of this representation in greater detail.

\subsection{Semantic Textual Similarity}
Semantic Textual Similarity (STS) is a benchmark that evaluates how similar two sentences are two each other, often on a scale from 0.0 - 5.0, with 0.0 representing entirely dissimilar sentences, and 5.0 representing semantically identical sentences (although not necessarily lexically exactly the same).

One benefit STS systems can offer in the context of IR is the ability to resolve synonyms or semantic similarities present in medical jargon, clinical symptoms, or the like. For instance, two terms such as "migraine" and "headache", or "abdominal pain" and "stomachache", can be semantically very similar. However, because these terms are not lexically identical it can be possible for systems to miss these. Although techniques such as query expansion exist in traditional IR systems, they are not always effective in identifying \emph{all} related words and consequently can miss particular associations that may semantically exist. STS systems can help resolve these similarities and identify terms that could be potentially relevant to the query which otherwise may have been missed.

As a concrete example to demonstrate the effectiveness of STS systems in practical IR tasks, we take the following passage, excerpted from \cite{zimran2016long}:

\begin{displayquote}
"Long-term efficacy and safety results of taliglucerase alfa up to 36 months in adult treatment-naive patients with Gaucher disease. Taliglucerase alfa is an intravenous enzyme replacement therapy approved for treatment of type 1 Gaucher disease (GD), and is the first available plant cell-expressed recombinant therapeutic protein. Herein, we report long-term safety and efficacy results of taliglucerase alfa in treatment-naive adult patients with GD. Patients were randomized to receive taliglucerase alfa 30 or 60 U/kg every other week, and 23 patients completed 36 months of treatment... All treatment-related adverse events were mild to moderate in intensity and transient. The most common adverse events were nasopharyngitis, arthralgia, upper respiratory tract infection, headache, pain in extremity, and hypertension...."
\end{displayquote}

\textit{The abstract has been truncated for brevity.}

\begin{figure}[h]
\includegraphics[width=15cm]{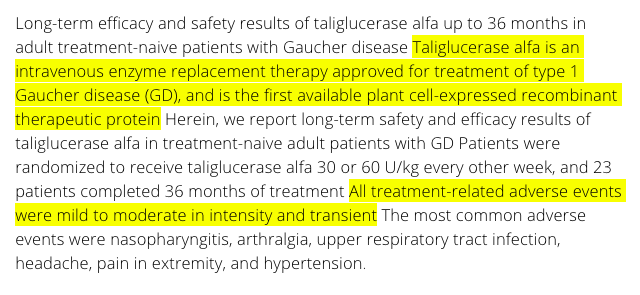}
\caption[Top 3 Ranked Sentences When Using Sentence Relevance Module.]{Top 3 Ranked Sentences when using Sentence Relevance module.}
\label{fig:migraine-example-alpha1}
\centering
\end{figure}

Given the above passage, and the following query:

\begin{displayquote}
"Can Gaucher disease treatment cause migraines?"
\end{displayquote}

a user would expect an IR system to return the last sentence in the excerpt, due to the fact that it mentions a connection between a potential treatment for Gaucher disease and headaches, and the semantic similarities between "migraines" and "headaches". However, when computing Sentence Ranking scores using the Binary Sentence Relevance model, we receive the highlighted sentences displayed in Figure \ref{fig:migraine-example-alpha1} as the Top 3 ranked sentences. Thus, it is evident the Sentence Relevance-based module did not consider the potential semantic similarity between the two terms to be significant enough to warrant a high ranking score.

\begin{figure}[h]
\includegraphics[width=15cm]{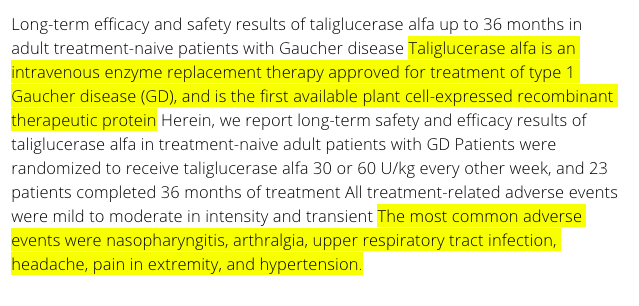}
\caption[Top 3 Ranked Sentences When Using Semantic Textual Similarity Module.]{Top 3 Ranked Sentences when using Semantic Textual Similarity module.}
\label{fig:migraine-example-alpha0}
\centering
\end{figure}

On the other hand, Figure \ref{fig:migraine-example-alpha0} shows the Top 3 sentences when using the STS module to generate sentence rankings. In this case, the particular sentence recieved a high enough score to warrant a Top 3 ranking.

This example seeks to demonstrate the value of an STS system in a practical IR situation. As will be demonstrated in subsequent Sections and Chapters, utilizing solely an STS based system is not effective in a Sentence Ranking task; thus, there needs to be an integration of these multiple modules.

\subsection{Semantic Information Availability}

\begin{table}[]
\emph{Given a query Q and a sentence S:}
\\
\begin{tabular}{|l|l|}
\hline
\textbf{Value} & \textbf{Description} \\ \hline
4 & Sentence S has the exact information related to query Q \\ \hline
3 & Sentence S has almost exact information related to query Q \\ \hline
2 & Sentence S has partial information related to query Q \\ \hline
1 & Sentence S has very little information related to query Q \\ \hline
0 & Sentence S has no information related to query Q \\ \hline
\end{tabular}
\caption[The Semantic Information Availability (SIA) Scale]{The Semantic Information Availability (SIA) scale}
\label{tab:sia-scale}
\end{table}

Existing scales used in NLP like Natural Language Inference (NLI) and Semantic Textual Similarity (STS) can be useful for finding and evaluting knowledge, like discussed previously. However, there is no standardized scale for information-seeking tasks that can answer the question \emph{Given a query Q and sentence S, how much information does S have that is needed to answer Q?} The motivation behind proposing a new scale, termed Semantic Information Availability (SIA), is to have a formal metric that can do so. Being able to evaluate query-sentence pairs in this manner can have direct practical benefits for tasks like IR, for which existing methodologies, as discussed previously, are not able to fully capture the necessities for effective ranking. 

The SIA scale is defined in Table \ref{tab:sia-scale}. A more detailed interpretation of the SIA scale is as follows:

\noindent \emph{Given question Q and sentence S:} \\
\textbf{4} - S can completely and unambiguously answer Q with no additional reasoning necessary \\
\textbf{3} - S can answer Q if some slight reasoning is applied to the information given in S (meaning the answer is not 100\% explicitly spelled out in S) \\
\textbf{2} - One other piece of information is needed to “link” the information given in S to answer Q (one-hop reasoning) \\
\textbf{1} - There is some information in S which, when reasoning is applied, can relate to part of the question Q. In other words, applying reasoning to S will get you to half or less of the full answer. In other words -- S requires multi (n $>$ 1)-hop reasoning to answer Q \\ 
\textbf{0} - There is absolutely nothing in S that can be used to answer any portion of Q, even if additional reasoning is applied \\

As can be observed from the definition of the SIA scale, as well as the definitions of each of the components in the scale, this task is very aligned with the high-level goals of Information Retrieval -- to be able to score how well a particular sentence answers a query. Accordingly, the intuition is that having a module which, given a query and sentence as an input, predicts a score from 0 to 4 reflecting the amount of information that the sentence has to answer the query will be very useful for this Semantic IR task.

The major challenge in being able to utilize the SIA scale as part of a model is that there is no existing dataset. Consequently, it is necessary to create very large amounts of data (in the tens of thousands of entries) in order to train a system which can be an accurate predictor for SIA. In order to rapidly generate the data necessary to train a system on the SIA task, a largely automated methodology was taken that utilized existing datasets and generative models.

First, specifically, automatic generation was used to come up with \texttt{(query, sentence)} pairs of Categories 0, 2, and 4, in the following process:

The Question Answering via Sentence Composition (QASC) Dataset was used as a base for science-related SIA \citep{khot2019qasc}. This dataset consists of 9,980 8-way multiple-choice questions about grade school science, and comes with a corpus of 17M sentences. Each question is annotated with two facts from the corpus that can be combined together to arrive at the answer \texttt{(question, possible\_answers, correct\_answer, fact1, fact2, combinedFact)}. The following rules were used to convert this dataset into an SIA dataset:  

\textbf{For Category 4 Samples:} For each gold row in QASC, create SIA entry \texttt{(Q=question, S=combinedFact)}.

\textbf{For Category 2 Samples:} For each gold row, there are two potential SIA entries: \texttt{(Q=question, S=fact1)} and \texttt{(Q=question, S=fact2)}. However, because this is a multiple-choice task and SIA is meant to be an open-ended dataset, one of the answers will not be suitable for the SIA task as it will exploit the multiple-choice nature of the QASC dataset due to the data generation templates used for QASC. Thus, to select the best sentence, the BERT Sentence Relevance model finetuned on the MS MARCO dataset (described earlier this Chapter) was applied on both pairs, and the higher scoring pair was selected. This method was determined to be a useable heuristic by randomly hand-annotating around 50 samples and recognizing that this method correctly selected the useable sentence for each sample. 

\textbf{For Category 0 Samples:} Using a RoBERTa \citep{liu2019roberta} STS model trained on the GLUE STS-B general English task, two gold rows were picked in the QASC dataset, where the STS score between the two gold questions was above a particular threshold, determined heuristically. As long as the two queries did not match, two SIA rows were created: \texttt{(Q=question1, S=combinedFact2)} and \texttt{(Q=question2, S=combinedFact1)}

Using this methodology, a preliminary SIA dataset in the science domain was created, consisting of roughly 30,000 rows of $(query, sentence, sia\_score)$ tuples.

From this initial dataset, two generative models are trained. Specifically, the OpenAI Generative Pretrained Transformer 2 (GPT-2) model is used \citep{radford2019languagegpt}. A GPT-2 Medium (355 Million parameters) is trained to generate a $query$ given a $sentence\_ 4$ value, and another is trained to generate a $sentence\_2$ given a $sentence\_4$. Both these models are trained only on the initially-generated dataset. These generative models enable creation of SIA samples in a purely unsupervised manner, rather than having to rely on existing datasets and repurposing them for the SIA task.

To generate Sentence 3 and Sentence 1 entries, the following method was used. For each Sentence 4, the ScispaCy library \citep{Neumann2019ScispaCyFA}, built on the SpaCy library \citep{spacy2}, was used to identify relevant biomedical entities from the sentence. Then, with a random probability, the entity is selected (and split into individual words if necessary). For each word, a substitution is found by finding the top 5 words by cosine similarity in a word2vec model trained on several clinical datasets, from which a random one is selected. The original word is swapped with its substitution. In this manner, a sentence from which part of the key entities are replaced with similar words is generated, and taken as Sentence 3 value. The same process is used to generate Sentence 1 entries; however, instead of using a Sentence 4 as the "base", the Sentence 2 is used.

Using both the generative models and word swapping techniques, a SIA dataset is constructed using sentences from the BioASQ training dataset. Roughly 200,000 SIA samples are created through this process. These samples are added to the first phase of auto-generated examples via the QASC dataset.

While the automatically-generated samples are comparatively less nuanced, and in some cases, less varied than examples which could be human-annotated, having an automated pipeline allows for the generation of the many examples needed to train a system from scratch.

Once the SIA dataset has been generated, a BERT-Large model is finetuned on this task, formulated as a regression task. The Mean Squared Error loss is used during the training of this model. During inference, a query and sentence are passed in to the model, which returns a score between 0.0 and 4.0, which is taken as the "SIA Score" for that particular query-sentence pair.

For application of the SIA module to the BioASQ Challenge, two versions of the model were developed; one using only the initial auto-generated data from QASC, and one using the data generated with the methodology described above. In practice, the former model was found to yield better results in the BioASQ Challenge.
\chapter{MULTI-PERSPECTIVE INFORMATION RETRIEVAL}
We have now discussed the semantics and methods behind different ways of evaluating and measuring query-sentence pairs. We have also noted the advantages and disadvantages of each of the methods with respect to their application to the IR task. We will now discuss the "fusion" component of the proposed system illustrated in Figure \ref{fig:multiperspective-1}. In order to determine how to most effectively combine this disparate models into a single ranking score, 7 broad methodologies were tried out and evaluated empirically on the BioASQ challenge. In this subsection, we will discuss these methodologies in detail and discuss their successes and failures, and describe the eventual system used. 

As described previously, Sentence Ranking was the primary focus of this system. We will therefore discuss the integration of the various components discussed in the previous Chapter into a single Sentence Ranking module. Next, we will discuss how the Sentence Ranking module was used in the formulation of a Document Retrieval module for the BioASQ Challenge.

\section{Component Fusion for Sentence Ranking}
The 7 broad methodologies used can be classified into two general categories, which we refer to as "Feedforward Fusion" and "Weighted Sum". We will divide the discussion into these two broad categories and detail the different efforts within each of these categories.

\subsection{Feedforward Fusion}
The main objective of the Feedforward Fusion method was to learn how to combine the different modules together into a single sentence ranking score. There were several approaches that were evaluated in order to attempt to learn this fusion methodology; the high-level approach of the Feedforward Fusion method is illustrated in Figure \ref{fig:feedforward-fusion}. There were two main ideas behind the Learnable Task, using binary Sentence Relevance as the learnable task, or comparing two sentences as a time to predict the higher-ranked one. Both these ideas and relevant approaches will be discussed.

\begin{figure}[h]
\includegraphics[width=12cm]{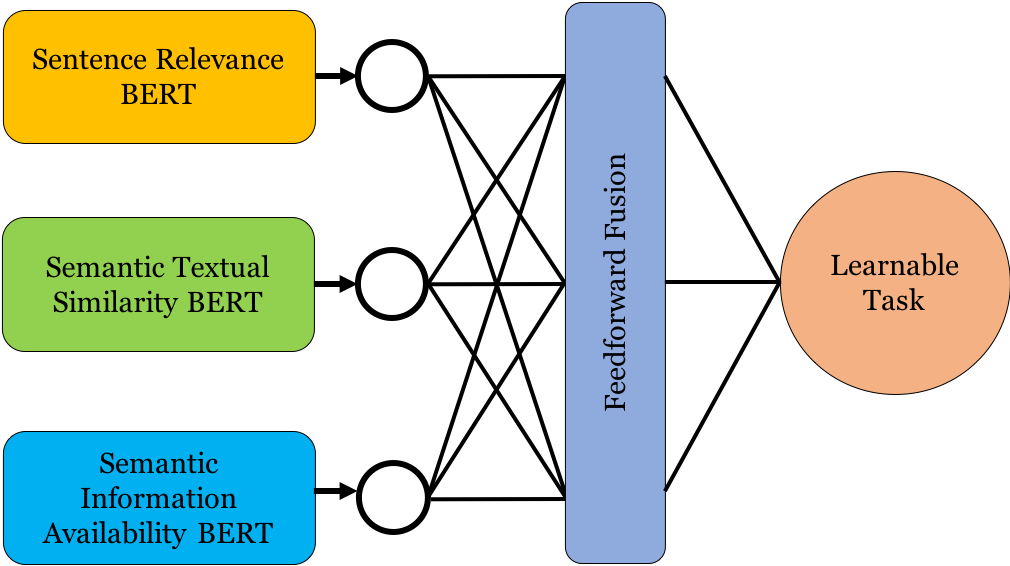}
\caption[General Architecture of Feedforward Fusion Method]{General architecture of Feedforward Fusion method}
\label{fig:feedforward-fusion}
\centering
\end{figure}

\subsubsection{Binary Sentence Relevance as Learnable Task}
The intuition behind using binary Sentence Relevance as the task to learn the fusion methodology was as follows:

\begin{itemize}
    \item Out of the three modules, binary Sentence Relevance is semantically the closest to the Sentence Ranking objective for IR.
    \item The best finetuned binary Sentence Relevance model only reaches an F1 score of 0.6066, indicating there is potential room for improvement in this task, possibly with the integration of the other modules. Furthermore, we have seen a correlation between higher performance on the Sentence Relevance task and higher performance on the BioASQ Sentence Ranking task, so there is value in improving the Sentence Relevance task.
\end{itemize}


\begin{figure}[h]
\includegraphics[width=14cm]{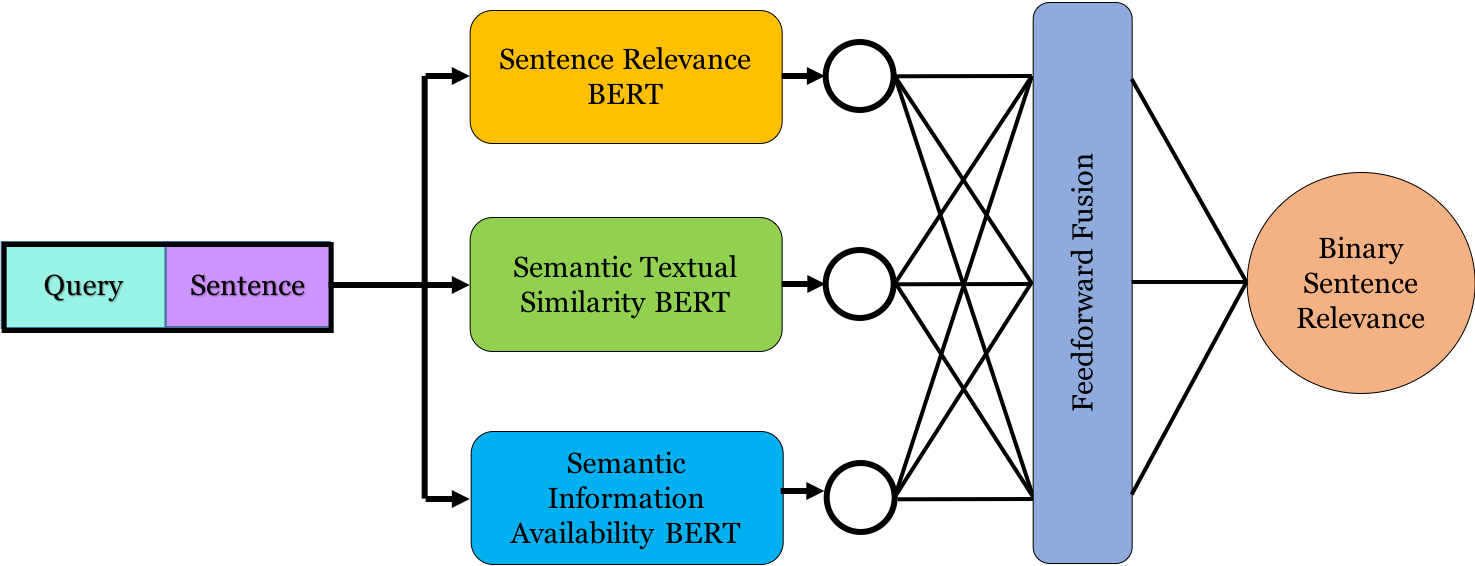}
\caption[Architecture for the Binary Sentence Relevance Feedforward Fusion Method]{Architecture for the Binary Sentence Relevance Feedforward Fusion method.}
\label{fig:feedforward-fusion}
\centering
\end{figure}
A variety of architectures were evaluated on their effectiveness to learn with teh Sentence Ranking objective. In the first model, the \texttt{(query, sentence)} pairs are passed through Sentence Relevance and STS BERT models, and the \texttt{CLS} tokens of the BERT models are passed through a Linear layer, which outputs a binary prediction. Other architectures consisted of the Sentence Relevance \texttt{CLS} token and STS prediction passed through the Linear Layer and both Sentence Relevance and STS predictions (rather than \texttt{CLS} tokens) passed through the Linear layer.

It was found experimentally that none of these methods resulted in a model that was able to perform better on the binary Sentence Relevance task compared to the Sentence Relevance-only model, and converged either at the same performance or lower performance. This indicates that, for the binary Sentence Relevance task in and of itself, the introduction of multiple perspectives is not valuable.

\subsubsection{Two Sentence Ranking as Learnable Task}
The intuition behind this approach was to train models to try to learn to compare two sentences, and determine what \emph{the ranking of Sentence 1 was respective to the ranking of Sentence 2, given a query}. To do this, two sentences were passed into the model with the goal to make a binary prediction on whether Sentence 1 ranked higher than Sentence 2. During inference time, all permutations of potential sentences would be fed through the model, and the scores for each Sentence 1 would be summed up to generate the Sentence Ranking score. The combinations of passing CLS layers and numerical predictions were also employed here.

\begin{figure}[h]
\includegraphics[width=15cm]{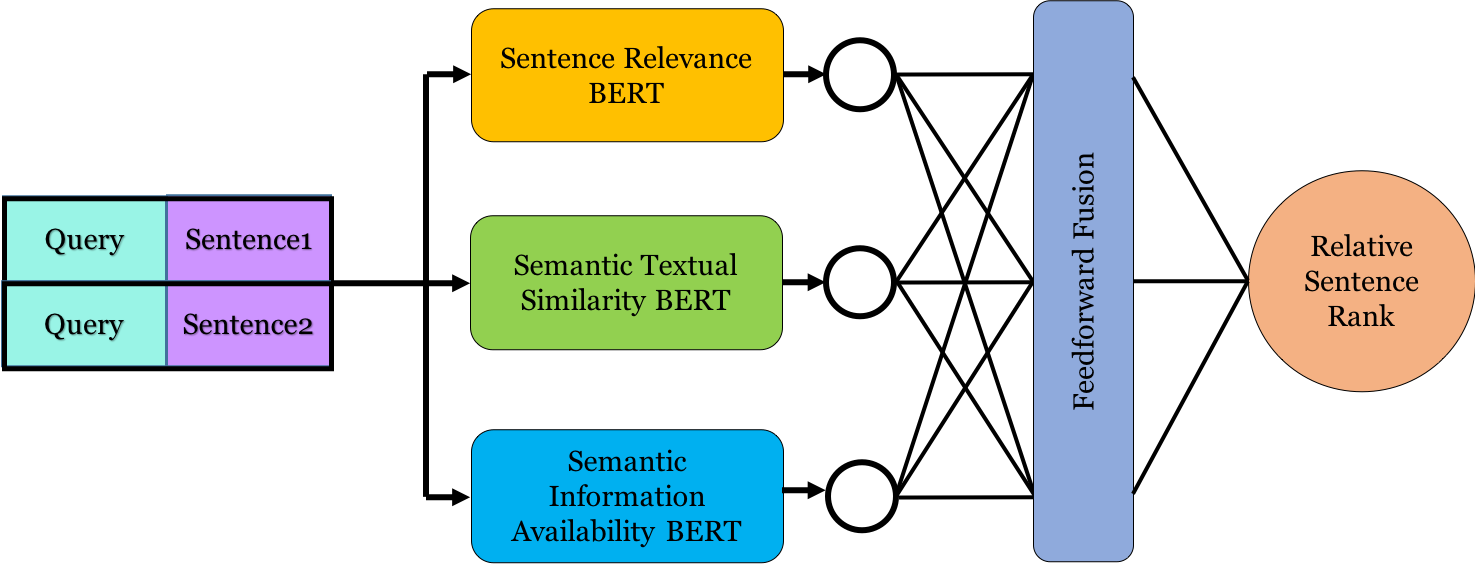}
\caption[Architecture for the Two Sentence Ranking Feedforward Fusion Method.]{Architecture for the Two Sentence Ranking Feedforward Fusion method.}
\label{fig:feedforward-fusion}
\centering
\end{figure}

The results of this methodology showed \emph{decreased} performance on the BioASQ Sentence Ranking objective compared to the performance of the "baseline" Sentence Ranking only-based model. This is likely because the objective difference between a sentence of Rank $n$ and a sentence of Rank $n+1$ in the BioASQ dataset is so subtle it is difficult to learn a method to distinguish the two. Moreover, after training this model and testing on the BioASQ Sentence Relevance challenge, the results significantly underperformed that of the baseline methodology. Parenthetically, this difficulty in distinguishing between particular biomedical data points is not altogether unexpected; similar challenges were encountered in our research when working on biomedical STS tasks. 

\subsubsection{Summary and Analysis}
All in all, neither of methods tested in the Feedforward Fusion category seemed to show any results that were more promising than the baseline Sentence Relevance model-only methodology. There are a few likely reasons why the Feedforward Fusion class of approaches did not work:

\begin{itemize}
    \item When trying to learn the fusion layer using a binary Sentence Relevance task, the components besides the Sentence Relevance component are not useful to generation of a prediction.
    \item Using a Relative Sentence Ranking task (comparing two Query-Sentence pairs and trying to predict which of is more relevant) is very difficult to learn, given that two sentences of Rank \textit{n} and Rank \textit{n+1} are very similar in terms of relevance.
\end{itemize}

Consequently, there may be a better means to formulate the learning task used in the Feedforward Approach, which we will leave for consideration in future work.

The second point yields an intuition that is very helpful in clarifying the next approach: \emph{Rather than considering "relevance" as an inherent semantic attribute of a sentence given a query, we can think of relevance as the amount of "evidence" that can be compiled about a sentence given a query that supports its usefulness in answering the query.} We will pursue this intuition and formulate it into the Weighted Sum technique, which we will now discuss.

\subsection{Weighted Sum}
The second broad methodology tried was that of the Weighted Sum. In this methodology, the Sentence Ranking score would be formulated as:

\begin{center}
    $S_{SENT} = \alpha_{1}S_{SentRelevance} + \alpha_{2}S_{STS} + \alpha_{3}S_{SIA}$
\end{center}


where $\alpha$ values were hyperparameters which would be learned over the development set. The intuition behind this approach was that, because learning the correct means to fuse together the different components were not effective as evidenced by the results of the Feedforward Fusion experiments, a potentially better way of determining how much to consider each component when fusing was to empirically see what combinations worked the best. 

This methodology was comparatively simpler than the Feedforward Fusion experiments. Results from each of the models were cached for efficiency, and hyperparameter combinations for all $\alpha$ values, between 0.0 and 1.0, at intervals of 0.1, were tested. The combination that produced the highest Snippet Mean Average Precision (the official metric for the BioASQ Sentence Ranking challenge) on the validation dataset was kept.

Initially, these results were tuned using the gold document lists from the BioASQ Challenge (meaning the values were optimized for the \textbf{maximum Sentence MAP score given the gold documents}.

However, evaluating on the BioASQ Sentence Ranking challenge also requires a Document Retrieval module to retrieve the top $n$ documents, where $n$ is up to 10, and then ranking the sentences from those retrieved documents. Thus, having a Document Retrieval module was necessary to evaluate the system on this particular challenge; moreover, it was possible the hyperparameters obtained through validating with gold documents were not the ideal hyperparameters in a setting where the correctness of the documents is not 100\%.  

We will therefore discuss the formulation of the Sentence Ranking task for the Document Retrieval task, and the Alternating Optimization algorithm used to tune the system, in the subsequent Section.

\section{Document Retrieval}
As discussed earlier, the main focus of this research was to develop a Sentence Ranking system, as it could be evaluated in the BioASQ Challenge, as well as utilized for the Medic Interface and Living Systematic Review tasks. Therefore, we choose to formulate the Document Retrieval task as (partially) a Sentence Ranking task in order to be able to use the same components we have developed earlier.

Specifically, we will formulate the Document Ranking task as a weighted sum with the "baseline" BM25 algorithm, and the Top 3 sentence scores of a particular document, as follows:

\begin{center}
    $S_{DOC} =( \alpha_{1} * S_{BM25}) +( \alpha_{2} * \sum_{i=0}^{3}(w_i * S_{SENT\_i}))$
\end{center}

where $\alpha$ and $w$ are hyperparameters that are learned over the validation set. Intuitively, this means that a document will be ranked using a combination of the bag-of-words-based BM-25 algorithm and the its Top 3 most relevant sentences (which are determined through the Semantic Sentence Ranking methodology discussed above).

\section{Semantic Document + Sentence Retrieval}

\begin{figure}[h]
\includegraphics[width=12cm]{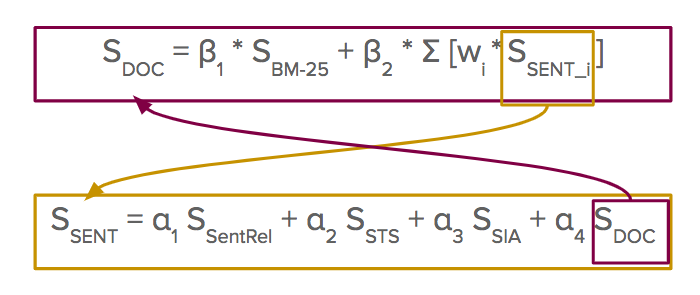}
\caption[Alternating Optimization Formulation of Document and Sentence Ranking.]{Alternating Optimization formulation of Document and Sentence Ranking.}
\label{fig:em-formulation}
\centering
\end{figure}

We will now take this formulation one step further by integrating the Document Ranking into the Sentence Ranking formulation as well. Doing so allows us to formulate an objective which we can represent as an Expectation Maximization (EM)-like algorithm (which we will term as an "Alternating Optimization" algorithm) described in Figure \ref{fig:em-formulation}.

This formulation captures some of the following intuitions:

\begin{itemize}
    \item A document's relevance can be captured (in part) by the relevance of its top $n$ sentences. This is an intuition that has also been followed in \citep{nogueira2019passage}.
    \item A sentence's relevance can be captured (in part) by the overall relevance of the document it is a part of.
\end{itemize}

There is an interesting recursive relationship between these two points -- that is, improving the Document Ranking could theoretically boost the Sentence Ranking performance; in turn, improving the Sentence Ranking could boost the Document Ranking performance. This sets up a recursive training loop which can be formulated in a similar manner as an Alternating Optimization algorithm. The overall algorithm to maximize the Sentence Mean Average Precision score, the primary benchmark for the BioASQ Sentence Ranking challenge, is detailed in Algorithm \ref{algo:EM-algorithm}.

\begin{algorithm}
\begin{algorithmic}
\STATE Given $S_{DOC}(\alpha, \beta, w) = \beta_1 * S_{BM-25} + \beta_2 * \sum_{i=1}^{3}(w_i * S_{SENT\_i}(\alpha, \beta, w))$
\STATE Given $S_{SENT}(\alpha, \beta, w) = \alpha_1 * S_{SentRelevance} + \alpha_2 * S_{STS} + \alpha_3 * S_{DOC}(\alpha, \beta, w) $ 
\STATE $\alpha \leftarrow [rand(0,1), rand(0,1), rand(0,1)]$
\STATE $\beta \leftarrow [rand(0,1), rand(0,1)]$
\STATE $w \leftarrow [rand(0,1), rand(0,1), rand(0,1)]$

\STATE $SENT\_MAP \leftarrow 0.0$
\STATE $DOC\_MAP \leftarrow 0.0$

\STATE $NEW\_SENT\_MAP \leftarrow S_{SENT}(\alpha, \beta, w)$
\STATE $NEW\_DOC\_MAP \leftarrow S_{DOC}(\alpha, \beta, w)$

\WHILE{$ NEW\_SENT\_MAP > SENT\_MAP$}

\STATE \textbf{E Phase}
\STATE $DOC\_MAP \leftarrow NEW\_DOC\_MAP$ 
\STATE $new\_\beta, new\_w \leftarrow maximize DOC\_MAP(\alpha, \beta, w)\ via\ HyperOpt(\beta, w)$
\STATE $\beta \leftarrow new\_\beta$
\STATE $w \leftarrow new\_w$
\STATE $NEW\_DOC\_MAP \leftarrow S_{DOC}(\alpha, \beta, w)$

\STATE \textbf{M Phase}
\STATE $SENT\_MAP \leftarrow NEW\_SENT\_MAP$ 
\STATE $new\_\alpha \leftarrow maximize SENT\_MAP(\alpha, \beta, w)\ via\ HyperOpt(\alpha)$
\STATE $\alpha \leftarrow new\_\alpha$
\STATE $NEW\_SENT\_MAP \leftarrow S_{SENT}(\alpha, \beta, w)$

\ENDWHILE
\end{algorithmic}
\caption[Algorithm for Maximizing Sentence Mean Average Precision Score using Alternating Optimization Methodology.]{Algorithm for maximizing Sentence Mean Average Precision score using Alternating Optimization methodology.}
\label{algo:EM-algorithm}
\end{algorithm}

Through running multiple iterations of the EM algorithm with randomly initialized parameters, some heuristics were identified to help initialize the parameters with values that would produce higher results. Primarily, initializing balanced $\alpha$, $\beta$, and $w$ values (i.e. making all values in the array for a particular parameter the same values) was found to be beneficial.

Moreover, conducting a hyperparameter search via gridsearch from a range of 0.0 to 1.0 by tenths (0.1) for each parameter was found to be computationally-expensive and time-consuming, even with caching of the Anserini and BERT model results. To circumvent this, two techniques were tried. First, gridsearch was used to conduct hyperparameter optimization, but with steps of 0.2. Once high-performing parameters were found, the algorithm was run a second time, with smaller steps around a smaller window around these parameters. The second technique was to utilize Bayesian Optimization using the BayesianOptimization Python package \citep{bayesianopt} over the parameters rather than gridsearch, to cover the search space with less computation. This second approach proved to be empirically far more successful in finding parameters that would maximize the Sentence MAP score on the validation set. When using Bayesian Optimization, the initial starting parameters were not needed.

Some variants of this methodology were attempted. The first was coming up with two sets of parameters, one set obtained through running the EM algorithm with the maximization objective of maximizing Sentence MAP, and the other set obtained through running the EM algorithm with the maximization objective of maximizing Document MAP scores. The next was the intuition that, although Sentence Ranking is a downstream task to Document Retrieval, maximizing the Document Retrieval F1 score is more important in improving the Sentence Ranking MAP score than maximizing the Document Retrieval MAP score. This notion will be further discussed in the following Chapter. Thus, in one variant, the Expectation step seeked to maximize F1, while the Maximization step seeked to maximize MAP.


This algorithm was run on the development set to obtain final values for $\alpha$, $\beta$, and $w$. During the test phase of the BioASQ Document and Sentence Ranking Challenge, these parameters are then used with the $S_{DOC}$ function to first fetch the top 10 Documents, and then with the $S_{SENT}$ function to fetch the top 10 Sentences from the top documents. In the next Chapter we will discuss evaluation results of this system on the BioASQ Challenge for Document Ranking and Snippet (Sentence) Ranking.

Table \ref{tab:hyperopt1-golddoc} lists the hyperparameters that were obtained following the Alternating Optimization algorithm for maximizing the Sentence Mean Average Precision metric.

The size of the $w$ vector represents how many scores of the top sentences from a particular document would be used in the Document Score formulation. For instance, having a $w$ vector of size 5 would mean the Top 5 sentence scores from a particular document would be weighted and utilized when ranking documents. However, it was found that when utilizing any number of sentences beyond the Top 3, the remaining sentences were not used. Even when using the Top 3 sentences, the majority of the weights were placed on the highest and second-highest sentences, while the 3rd highest-scored sentence was not utilized. These findings are reflected in Table \ref{tab:hyperopt1-golddoc}.

\begin{table}[t]
\centering
\begin{tabular}{|c|c|}
\hline
\textbf{Hyperparameter} & \textbf{Value} \\ \hline
$\alpha_1$ Sentence Relevance & 0.6123 \\ \hline
$\alpha_2$ (Semantic Textual Similarity) & 0.2664 \\ \hline
$\alpha_3$ (Semantic Information Availability) & 0.0785 \\ \hline
$\alpha_4$ (Document Score) & 0.9879 \\ \hline
$\beta_1$ (BM-25 Score) & 0.0002 \\ \hline
$\beta_2$ (Sentence Relevance Score) & 0.8523 \\ \hline
$w_1$ (Top Sentence Score 1) & 0.9938 \\ \hline
$w_2$ (Top Sentence Score 2) & 0.0338 \\ \hline
$w_3$ (Top Sentence Score 3) & 0.0271 \\ \hline
\end{tabular}
\caption[Learned Hyperparameter Values Over BioASQ Sentence Relevance Development Dataset, with Gold Documents.]{Learned hyperparameter values over BioASQ Sentence Relevance development dataset, with gold Documents.}
\label{tab:hyperopt1-golddoc}
\end{table}

\chapter{BIOASQ DOCUMENT AND SENTENCE RANKING CHALLENGE}
The system described in this work was initially evaluated on the BioASQ 6B Phase A Document and Snippet ranking challenges, due to the comparatively larger amount of existing literature available regarding this challenge for comparison and analysis purposes. However, after evaluation on the BioASQ 6B results, this system was also evaluated on the BioASQ 7B Phase A Document and Snippet ranking challenges. Note that in the BioASQ Snippet Ranking component, a "snippet" is not necessarily identical to a "sentence" (it can be a portion of a sentence or multiple sentences); however, practically speaking, the majority of snippets were single sentences. Thus, for the purposes of this system, a snippet is considered equivalent to a single sentence.

This Chapter will cover the performance of this system, as well as some other systems, on the BioASQ Challenges, along with analysis of the results. Given that in the BioASQ Challenge, Snippet Ranking is a downstream task of Document Ranking, we will discuss the Document Ranking results first.

\begin{figure}[t]
\includegraphics[width=16cm]{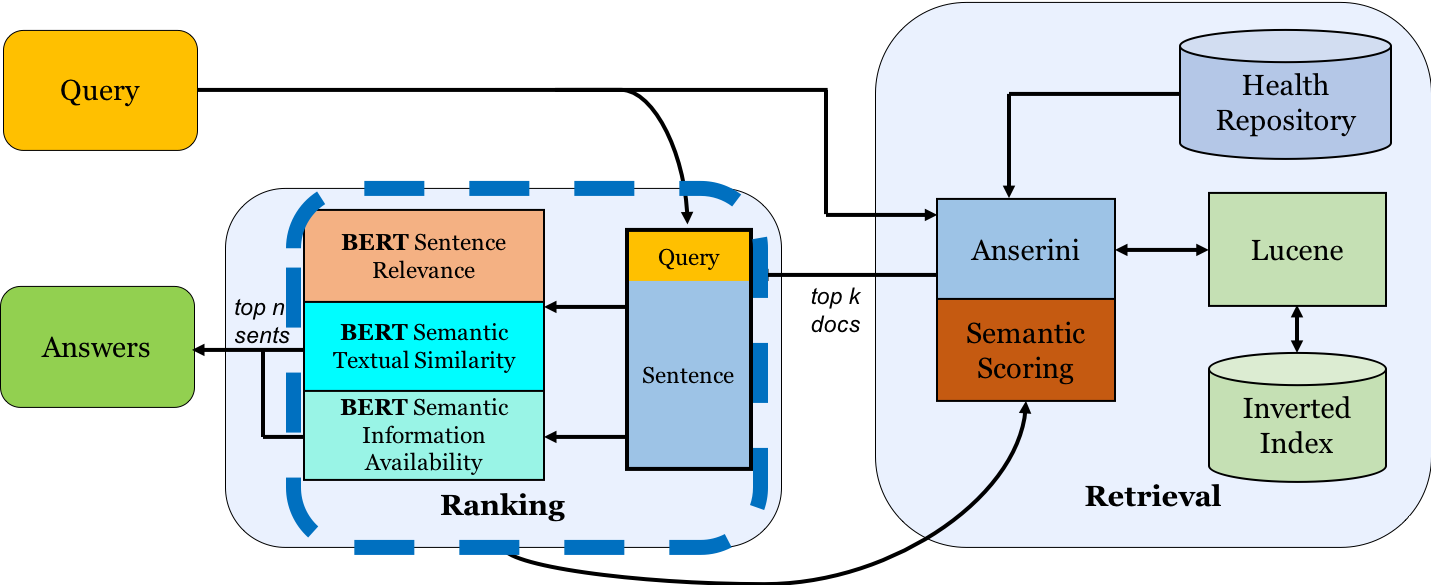}
\caption[Overall Semantic IR System Architecture, with Document Ranking Formulated as a Sentence Ranking Task]{Overall Semantic IR system architecture, with Document Ranking formulated as a Sentence Ranking task}
\label{fig:overallarchitecture-2}
\centering
\end{figure}

Figure \ref{fig:overallarchitecture-2} is the high-level overview of the system detailed in the previous Chapter applied to the BioASQ Challenge.

\emph{Note:} For both the BioASQ 6B and BioASQ 7B Phase A competitions, there were 5 "batches" of test data which teams participated in. The tables shown in this section are the BioASQ 6B Phase A, Batch 2; and BioASQ 7B Phase A, Batch 4 results. These results are generally representative of those from the other batches; the other batch results are available in the Appendix of this document.

\section{BioASQ Document Retrieval Challenge}
The objective for the BioASQ Document Retrieval Challenge is, given a query, to locate the top $n$ documents from the MEDLINE/PubMed Baseline repository for the particular query, where $n$ is between 0 and 10 documents. The BioASQ 6 Challenge is based on the MEDLINE/PubMed 2018 Baseline repository, which consists of about 25 million documents, while the BioASQ 7 Challenge is based on the MEDLINE/PubMed 2019 Baseline repository, which consists of about 29 million documents. A "document" in the context of the BioASQ Challenge and MEDLINE/PubMed repository is the title and abstract of all completed citations in MEDLINE present in that year.

As mentioned earlier, the system was intended to be evaluated on the BioASQ 6B Challenge and was subsequently also evaluated on the BioASQ 7B Challenge. As a result, the system was trained using the BioASQ 6B Training Set, rather than the BioASQ 7B Training Set, which contains additional data. Consequently, re-training and hyperparameter tuning the system for the BioASQ 7B Challenge specifically may lead to improved results in that task, which is left as an exercise for Future Work.

The results for the BioASQ 6B Phase A Challenge, alongside some other teams' submissions to the challenges, are detailed in Figure \ref{tab:bioasq6-doc}. The results for the BioASQ 7B Phase A Challenge are detailed in Figure \ref{tab:bioasq7-doc}. The \underline{Semantic-Doc} submission is the system that has been trained using the Alternating Optimization algorithm with the objective to \emph{maximize Document Mean Average Precision}; the \underline{Semantic-Sent} submission is the system that has been trained using the AO algorithm with the objective to \emph{maximize Sentence Mean Average Precision}. When using documents for the downstream Sentence Ranking task, although the Document Ranking score is used in the formulation of the Sentence Ranking score, the rank of the document with respect to the other documents is explicitly not utilized; the more important factor is that the document made the Top 10 list, and as few as possible irrelevant documents made the Top 10 list. This can be reflected in the fact that for both BioASQ Challenges, the Semantic-Sent system had a higher Document F1 score (which does not account for ordering and only on presence or absence of the document), while the Semantic-Doc system had a higher Document MAP score.

Further analysis will be discussed after the Sentence Retrieval scores have been mentioned.

\begin{table}[p]
\begin{tabular}{|c|c|c|c|
>{\columncolor[HTML]{FFFC9E}}c |c|}
\hline
\textbf{System} & \textbf{\begin{tabular}[c]{@{}c@{}}MPrec\\ Docs\end{tabular}} & \textbf{\begin{tabular}[c]{@{}c@{}}MRec\\ Docs\end{tabular}} & \textbf{\begin{tabular}[c]{@{}c@{}}F-Measure\\ Docs\end{tabular}} & \textbf{\begin{tabular}[c]{@{}c@{}}MAP\\ Docs\end{tabular}} & \textbf{\begin{tabular}[c]{@{}c@{}}GMAP\\ Docs\end{tabular}} \\ \hline
ustb\_prir3 & 0.3121 & 0.6379 & 0.3396 & 0.2512 & 0.0639 \\ \hline
ustb\_prir4 & 0.3121 & 0.6379 & 0.3396 & 0.2512 & 0.0639 \\ \hline
aueb-nlp-4 & 0.3220 & 0.6431 & 0.3479 & 0.2500 & 0.0660 \\ \hline
aueb-nlp-2 & 0.3210 & 0.6420 & 0.3475 & 0.2470 & 0.0701 \\ \hline
testtext & 0.3201 & 0.6355 & 0.3464 & 0.2467 & 0.0634 \\ \hline
ustb\_prir1 & 0.3201 & 0.6355 & 0.3464 & 0.2467 & 0.0634 \\ \hline
ustb\_prir2 & 0.3221 & 0.6618 & 0.3519 & 0.2458 & 0.0795 \\ \hline
aueb-nlp-3 & 0.3160 & 0.6365 & 0.3423 & 0.2416 & 0.0646 \\ \hline
aueb-nlp-1 & 0.3060 & 0.6294 & 0.3332 & 0.2319 & 0.0560 \\ \hline
\underline{Semantic IR} & \underline{0.2684} & \underline{0.5697} & \underline{0.2952} & \underline{0.2085} & \underline{0.0324} \\ \hline
\end{tabular}
\caption[BioASQ 6B Phase A Document Ranking Results]{BioASQ 6B Phase A Document Ranking results.}
\label{tab:bioasq6-doc}
\end{table}

\begin{table}[p]
\begin{tabular}{|c|c|c|c|
>{\columncolor[HTML]{FFFC9E}}c |c|}
\hline
\textbf{System} & \textbf{\begin{tabular}[c]{@{}c@{}}MPrec\\ Docs\end{tabular}} & \textbf{\begin{tabular}[c]{@{}c@{}}MRec\\ Docs\end{tabular}} & \textbf{\begin{tabular}[c]{@{}c@{}}F-Measure\\ Docs\end{tabular}} & \textbf{\begin{tabular}[c]{@{}c@{}}MAP\\ Docs\end{tabular}} & \textbf{\begin{tabular}[c]{@{}c@{}}GMAP\\ Docs\end{tabular}} \\ \hline
aueb-nlp-1 & 0.2541 & 0.6668 & 0.2998 & 0.2102 & 0.0316 \\ \hline
aueb-nlp-2 & 0.2531 & 0.6523 & 0.2992 & 0.2092 & 0.0279 \\ \hline
aueb-nlp-4 & 0.2481 & 0.6445 & 0.2948 & 0.2080 & 0.0268 \\ \hline
aueb-nlp-5 & 0.4537 & 0.6416 & 0.4580 & 0.1968 & 0.0291 \\ \hline
aueb-nlp-3 & 0.2401 & 0.6451 & 0.2857 & 0.1962 & 0.0282 \\ \hline
lh\_sys4 & 0.2230 & 0.6121 & 0.2695 & 0.1752 & 0.0186 \\ \hline
\underline{Semantic IR} & \underline{0.2170} & \underline{0.5867} & \underline{0.2592} & \underline{0.1777} & \underline{0.0176} \\ \hline
MindLab QA...  & 0.2080 & 0.5664 & 0.2463 & 0.1724 & 0.0121 \\ \hline
MindLab QA... & 0.2080 & 0.5664 & 0.2463 & 0.1724 & 0.0121 \\ \hline
MindLab QA...  & 0.2080 & 0.5664 & 0.2463 & 0.1724 & 0.0121 \\ \hline
\end{tabular}
\caption[BioASQ 7B Phase A Document Ranking Results]{BioASQ 7B Phase A Document Ranking results.}
\label{tab:bioasq7-doc}
\end{table}

\section{BioASQ Sentence Retrieval Challenge}
In the scope of the BioASQ Snippet Retrieval Challenge, the system must retrieve the top $n$ snippets from the top $k$ documents that have already been retrieved by the system, given a query. $n$ and $k$ are each integers between 0 and 10, and for the purposes of this system, a snippet is equivalent to a single sentence. Therefore, Snippet Retrieval is a downstream task of the Document Retrieval challenge; the more effective the Document Retrieval system is in identifying documents with relevant snippets, the more effective the Snippet Retrieval system will be.

\begin{table}[p]
\begin{tabular}{|c|c|c|c|
>{\columncolor[HTML]{FFFC9E}}c |c|}
\hline
\textbf{System} & \textbf{\begin{tabular}[c]{@{}c@{}}MPrec\\ Snippets\end{tabular}} & \textbf{\begin{tabular}[c]{@{}c@{}}MRec\\ Snippets\end{tabular}} & \textbf{\begin{tabular}[c]{@{}c@{}}F-Measure\\ Snippets\end{tabular}} & \textbf{\begin{tabular}[c]{@{}c@{}}MAP\\ Snippets\end{tabular}} & \textbf{\begin{tabular}[c]{@{}c@{}}GMAP\\ Snippets\end{tabular}} \\ \hline

\underline{Semantic IR} & \underline{0.3805} & \underline{0.3643} & \underline{0.3357} & \underline{0.3731} & \underline{0.0787} \\ \hline
aueb-nlp-5 & 0.3852 & 0.2976 & 0.2653 & 0.3187 & 0.0352 \\ \hline
MindLab QA...  & 0.2878 & 0.2307 & 0.1985 & 0.2736 & 0.0065 \\ \hline
MindLab QA...  & 0.2888 & 0.2298 & 0.1986 & 0.2695 & 0.0071 \\ \hline
MindLab QA... & 0.2888 & 0.2298 & 0.1986 & 0.2695 & 0.0071 \\ \hline
aueb-nlp-4 & 0.2873 & 0.2146 & 0.1850 & 0.2337 & 0.0231 \\ \hline
aueb-nlp-3 & 0.2746 & 0.2041 & 0.1749 & 0.2272 & 0.0210 \\ \hline
aueb-nlp-2 & 0.2768 & 0.2133 & 0.1826 & 0.2256 & 0.0236 \\ \hline
aueb-nlp-1 & 0.2716 & 0.2055 & 0.1749 & 0.2226 & 0.0202 \\ \hline
ustb\_prir4 & 0.2179 & 0.6188 & 0.2566 & 0.1731 & 0.0205 \\ \hline
\end{tabular}
\caption[BioASQ 6B Phase A Snippet Ranking Results.]{BioASQ 6B Phase A Snippet Ranking results.}
\label{tab:bioasq6-sent}
\end{table}

\begin{table}[p]
\begin{tabular}{|c|c|c|c|
>{\columncolor[HTML]{FFFC9E}}c |c|}
\hline
\textbf{System} & \textbf{\begin{tabular}[c]{@{}c@{}}MPrec\\ Snippets\end{tabular}} & \textbf{\begin{tabular}[c]{@{}c@{}}MRec\\ Snippets\end{tabular}} & \textbf{\begin{tabular}[c]{@{}c@{}}F-Measure\\ Snippets\end{tabular}} & \textbf{\begin{tabular}[c]{@{}c@{}}MAP\\ Snippets\end{tabular}} & \textbf{\begin{tabular}[c]{@{}c@{}}GMAP\\ Snippets\end{tabular}} \\ \hline
aueb-nlp-2 & 0.3254 & 0.4308 & 0.3048 & 0.3409 & 0.0344 \\ \hline
aueb-nlp-1 & 0.3209 & 0.4321 & 0.3018 & 0.3249 & 0.0281 \\ \hline
\underline{Semantic IR} & \underline{0.2959} & \underline{0.3414} & \underline{0.2568} & \underline{0.3030} & \underline{0.0151} \\ \hline
aueb-nlp-2 & 0.3254 & 0.4308 & 0.3048 & 0.3409 & 0.0344 \\ \hline
aueb-nlp-1 & 0.3209 & 0.4321 & 0.3018 & 0.3249 & 0.0281 \\ \hline
\underline{Semantic IR} & \underline{0.2959} & \underline{0.3414} & \underline{0.2568} & \underline{0.3030} & \underline{0.0151} \\ \hline
aueb-nlp-5 & 0.3256 & 0.4403 & 0.3010 & 0.2976 & 0.0379 \\ \hline
MindLab QA... & 0.2276 & 0.2857 & 0.2093 & 0.2214 & 0.0052 \\ \hline
aueb-nlp-3 & 0.2563 & 0.3581 & 0.2346 & 0.2213 & 0.0196 \\ \hline
aueb-nlp-4 & 0.2550 & 0.3325 & 0.2318 & 0.2173 & 0.0178 \\ \hline
MindLab Red... & 0.2168 & 0.2718 & 0.1982 & 0.2000 & 0.0067 \\ \hline
MindLab QA... & 0.2112 & 0.2317 & 0.1819 & 0.1931 & 0.0058 \\ \hline
MindLab QA... & 0.1998 & 0.2669 & 0.1865 & 0.1892 & 0.0064 \\ \hline
\end{tabular}
\caption[BioASQ 7B Phase A Snippet Ranking Results]{BioASQ 7B Phase A Snippet Ranking results.}
\label{tab:bioasq7-sent}
\end{table}

\clearpage
\section{Analysis of the BioASQ Competition Results}
In the Document Ranking task, the system generally placed below the Top 5 submissions in the BioASQ 6 and 7 Document Ranking challenges. However, in the Snippet Ranking task, the system generally placed around 1st and 4th on the BioASQ 6 and 7 Snippet Ranking challenges, respectively, in terms of systems and 1st and 2nd place, respectively, in terms of participants. The exact results are provided in the Appendix of this document.

The Sentence Ranking module seems to perform very well with respect to the upstream Document F1 score compared to the other systems. Specifically, the Document F1 scores for the two systems outperforming this system in Snippet MAP in the BioASQ 7 challenge are 0.3409 and 0.3249, respectively. This system's Document F1 score in that challenge is 0.2592. This seems to imply that, given a higher performance in the Document Ranking task, the Snippet Ranking performance may come closer or exceed the other systems'. The comparatively high performance of the Sentence Ranking module in BioASQ 6, where the difference between this system's F1 Document Score and that of other top-performing systems was considerably smaller than in BioASQ 7, appears to support this idea. All in all, the value of introducing Multiple Perspectives when conducting Semantic Ranking, as well as incorporating the scores of the top sentences from a document that a sentence belongs to into that particular sentence's Ranking score, seems to be empirically validated through the Snippet Ranking performance in the BioASQ Challenge.

There is potential to improve the Document Ranking module, which as noted above will likely have the downstream impact of boosting the Sentence Ranking performance as well. Although the main focus in this work was developing a Sentence Ranking module, and formulating the Document Ranking task as partially a Sentence Ranking task, focusing on developing a Document Ranking module with a greater emphasis on document-specific features may lead to improved performance in this area. 

A further discussion on the potential improvements of this system is left to the Future Work portion of this document.
\chapter{PRACTICAL APPLICATIONS OF THE SEMANTIC IR SYSTEM}

This work presents practical contributions to Information Retrieval in the biomedical/clinical domains. The two primary such contributions are:

\begin{enumerate}
    \item An interactive interface for medics to retrieve relevant information to a query from a clinical guidebook
    \item The application of the system to the Living Systematic Review task
\end{enumerate}

This Chapter will detail the technical work and contributions involved in the process of developing these software applications. It will also provide a holistic evaluation and reflection on this work, as well as potential use cases for these applications.

\section{Medic App}
The Medic App is an deployed instance of the Semantic IR system with an interactive front-end interface. This contribution directly ties the research detailed in this work to the practical use of such a system described in the Motivation section of this document. As a demonstration, an interactive version of the Semantic IR system has been implemented on the U.S Special Operations Forces Medical Handbook. The Medic App facilitates medical professionals or other users to quickly and automatically locate information from the handbook relevant to clinical query. This frees users from having to manually read through the medic handbook to locate the information necessary to address a particular clinical query and instead jump directly to the relevant information, allowing clinical sites with limited time or resources to spend more effort on medical care.

In this Section, we will discuss the technical components of the application and demonstrate the value of utilizing the Semantic IR system for interfacing with the handbook as opposed to using existing PDF software search.

The Medic App comprises of multiple different components that bridge the IR system to usable software. This Section will discuss the following components in detail:

\begin{itemize}
    \item Medic App User Interface
    \item Semantic IR Application Programming Interface (API)
\end{itemize}

\subsection{Medic App Interface}

The Medic App interface can be accessed via web browser. When a user visits the web page, a querying and section analysis page will be displayed. The interface allows the user to perform the following actions:

\begin{itemize}
    \item Input a query in natural language
    \item Adjust the number of relevant results to locate
    \item View the system's confidence in the returned results
    \item Search across various data repositories, including the medic handbook (default), PubMed Baseline 2018 and 2019 repositories, or import a specific PubMed document by its Document ID
\end{itemize}

\begin{figure}[t]
\includegraphics[width=15cm]{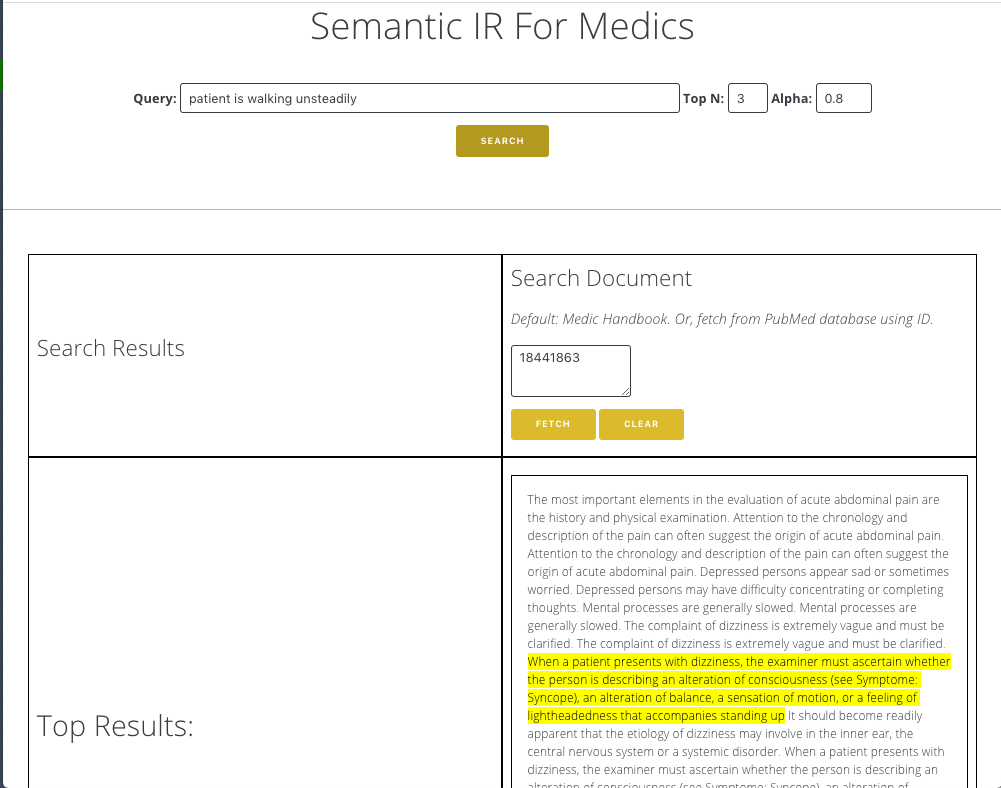}
\caption[Medic App User Interface]{Medic App user interface.}
\label{fig:medicdemo-1}
\centering
\end{figure}

The user inputs a query in the interface, and the system locates the most relevant sentences through the Medic Handbook and returns those results. The results are highlighted along with their adjacent sentences to provide better context to the results. Figure \ref{fig:medicdemo-1} is a screen capture of the primary interface to the IR module.

\subsection{Semantic IR API}
To facilitate real-time, custom user requests through the interface, a custom Application Programming Interface (API) was developed to serve requests to and from the IR system. Via the API, users can submit any query to the IR system, adjust parameters such as the number of relevant results to locate, and pass any custom data to perform IR across. The API is called via \texttt{GET} request in the following manner:

\begin{center}
    \texttt{http://<URL>/bert-ir?query=<QUERY>\&sentences=<TEXTDATA>\&
    \\ topn=<TOPN>\&alpha=<ALPHA>}
\end{center}

(where \texttt{URL} is the URL on which the API is served.)

The parameters are as follows:

\begin{itemize}
    \item \textbf{query}
    
    A query as a String. For example, \texttt{What happens during sceptic shock?}
    
    \item \textbf{sentences} (Optional)
    
    If the user chooses to search across custom text, the text can be passed through, and the system will search across that text. Otherwise, the default medic handbook data index will be used.
    
    \item \textbf{topn} (Optional)
    
    The number of top sentences to highlight and select for the user.
    
    \item \textbf{alpha} (Optional)
    
    The alpha values, as defined in the Sentence Ranking module formulation earlier in this document.
\end{itemize}

The code for hosting such an API is available on the public repository of this work. The specified API endpoints can be utilized to develop custom user interfaces across different devices. Although the neural network models can be hosted on CPU, deploying the API on CPU-only will result in retrieval times on the order of roughly 10 seconds (depending on the length of documents the ranking is conducted over), as opposed to 1-2 seconds if utilizing GPU.

Moreover, because the IR processing is done server-side, a user-facing tool can be deployed on low-powered devices if they are connected to the internet or a local intranet, such as smartphones or tablets. On the other hand, if an interface is deployed to a location with intermittent or limited internet access, the API can be deployed on a local machine so that all IR functionality will be available offline.

There are apparent advantages to the application of this system to the medic handbook with respect to the ability to find relevant semantic information that may not match a query textually. Some examples demonstrative of the capabilities of the system are included below:

\begin{itemize}
    \item
    \textbf{Query:} "patient is stumbling unsteadily"
    
    \textbf{Top Results:}
    \begin{itemize}
        \item Abnormal gait without dizziness is most likely ataxia (difficulty walking), a motor control problem.
        
        \item When a patient presents with dizziness, the examiner must ascertain whether the person is describing an alteration of consciousness (see Symptome: Syncope), an alteration of balance, a sensation of motion, or a feeling of lightheadedness that accompanies standing up.
    \end{itemize}
    
    \item
    \textbf{Query:} "patient not looking up"
    
    \textbf{Top Results:}
    \begin{itemize}
        \item A depressed person will often avoid eye contact, preferring to gaze downward.
    \end{itemize}
    
    \item
    \textbf{Query:} "What are causes of stomachache?"
    
    \textbf{Top Results:}
    \begin{itemize}
        \item Introduction: Acute abdominal pain is an internal response to a mechanical or chemical stimulus.
        
        \item Attention to the chronology and description of the pain can often suggest the origin of acute abdominal pain.
        
        \item The pain can be separated into three categories: visceral (dull and poorly characterized), somatoparietal (more intense and precisely localized) and referred (pain felt remote from the origin.
    \end{itemize}
    
\end{itemize}

The semantic nature of the IR system can be seen in the results from the examples above, in which there are no direct string matches. Instead, the system is able to resolve concepts like "stumbling unsteadily" and "dizziness" or "lightheadedness", and "not looking up" and "gaze downward". Moreover, the value of the STS component becomes evident in the third example. At the default alpha balance of 0.8 (which for this system means a 0.8 weight for Sentence Relevance module and 0.2 weight for STS module), only the first result from the third query is selected. However, after balancing the alpha a little more towards the STS module, the remaining two answers are also retrieved. The STS module is able to resolve two distinct but potentially similar clincal symptoms, "stomachache" and "abdominal pain" and bring a better user result.

\section{Living Systematic Review Task}

In this section we propose a challenge for the application of a Semantic Information Retrieval system to a clinical workflow. We will introduce the task, provide background information, and illustrates the workflow necessary to achieve the final objective. 

\subsection{Systematic Review}
A Systematic Review is a survey methodology used to evaluate and make note of new literature about a particular clinical topic and their clinical findings. Uman describes the goal of Systematic Reviews in "Systematic Reviews and Meta-Analyses" as "reducing bias by identifying, appraising, and synthesizing all relevant studies on a particular topic" \citep{uman2011systematic}. Carrying out Systematic Reviews is an essential task for consolidating the findings of dozens of studies into a single document, upon which further Meta-Analysis can be performed to be able to gain additional insights from the aggregation of literature. 

A Systematic Review is generally conducted by a team of domain experts, ranging from clinical librarians to physicians and assistants. The expertise and efforts of these different individuals is utilized at different stages of the Systematic Review pipeline.

\begin{figure}[t]
\includegraphics[width=15cm]{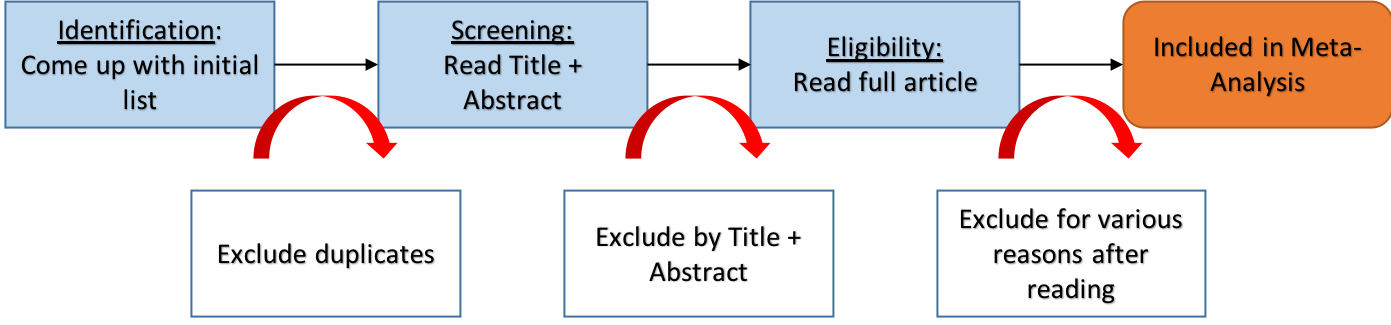}
\caption[Living Systematic Review Workflow]{Living Systematic Review workflow}
\label{fig:lsr-workflow}
\centering
\end{figure}

Although the exact workflow for conducting a Systematic Review can vary, typically, given a particular topic, a search strategy is devised by domain-level experts (such as a clinical librarian) to retrieve a candidate list of potentially relevant documents from multiple repositories of research papers, termed the "Identification" phase. Next, the candidate documents are manually filtered through a review of their titles and abstracts, and documents that appear to be irrelevant are excluded, termed the "Screening" phase. (One thing to note is that throughout the pipeline, documents are not selected to be "included" but instead always selected to be "excluded" or sent to the next phase in the pipeline, until the final list of documents remains.) After the initial screening is completed, the documents remaining in the pipeline are read more closely to determine whether they are relevant to the objective of the Systematic Review or not, termed the "Eligibility" phase. For example, items like the candidates in clinical trials, particular drugs evaluated, or study size are factors that determine whether a particular document remains in the pipeline or is removed. At the end of the Eligibility phase, the documents remaining in the pipeline are included in the resulting paper containing a summary and analysis of these results, called a Meta-Analysis document. This high level workflow is detailed in Figure \ref{fig:lsr-workflow}. Uman breaks down these steps into greater detail, listing 8 Stages of a Systematic Review and Meta-Analysis \citep{uman2011systematic}.

\subsection{Living Systematic Review}
A Living Systematic Review is envisioned to be a Systematic Review that is semi- or fully-automated, with the review process being aided by NLP algorithms. There is a strong motivation for this process -- conducting Systematic Reviews manually requires significant human investment, necessitating tens of hours of review. Moreover, due to the often fast pace of research in particular clinical areas, a Systematic Review can become out-of-date within months of its publication, requiring the process to be carried out every few months in order to ensure the latest information is present within these reviews.

Having a partially- or fully-automated Systematic Review process would enable it to be regularly updated without encountering as many issues of manual investment when trying to scale the process to be run more often. Consequently, such a system could become a "Living" Systematic Review, where the pipeline is run frequently enough to ensure the papers compiled as part of the Systematic Review are up-to-date in their findings. 

There are important considerations when seeking to automate parts or all of the Systematic Review process. First, the penalties for excluding a relevant document and including an irrelevant document are not identical, and need to be considered on a case-by-case basis. For instance, if the system is only automated in its initial steps and manually reviewed in subsequent steps, including an irrelevant document may not be a major problem, as the human reviewers can flag and exclude it downstream in the pipeline. However, if relevant documents are systematically excluded by the particular system, the manual reviews will never see those documents downstream in the pipeline, thereby enabling production of Systematic Reviews which may be overlooking potentially key studies or findings.

Another important consideration is the wide use-case for an automated system. Systematic Reviews are conducted across medical specialities and domains, and for a Living Systematic Review application to be broadly useful, the automated modules should be able to adapt to different domains and clinical areas with minimal fine-tuning required. 

Finally, Living Systematic Reviews are generally performed on constant intervals (for instance, every 3 months). An system should be able to learn from prior Systematic Reviews conducted, along with any human interventions or adjustments made throughout the course of the pipeline for the previous Systematic Reviews. Over time, the system should theoretically adapt in response to the human changes to minimize the amount of manual intervention that is necessary.

\subsection{Proposed Challenges for Living Systematic Reviews}
In this subsection, we will propose several challenges related to the integration of Semantic Information Retrieval and similar systems to the Living Systematic Review tasks. These proposals were developed through conversations with domain-level experts, including a clinical team that has conducted numerous Systematic Reviews and is working on developing a Living Systematic Review frontend and backend pipeline. 

As detailed earlier in this Section, a Systematic Review (and by extension, Living Systematic Review) pipeline generally consists of multiple stages, at which candidate documents are selected for exclusion or kept in the pipeline. The following proposed challenges will focus on developing systems around these various steps in the pipeline.

\subsubsection{Document Identification Task}

The first task is around the "Identification" phase. In this phase, domain-level experts like clinical librarians generally work alongside key stakeholders, like the physicians or researchers conducting these clinical reviews, to come up with search queries that will fetch potentially-relevant documents from a range of clinical research paper databases such as Embase, MEDLINE, and EBM Reviews - Cochrane Central Register of Controlled Trials. Depending on the complexity and breadth of the subject matter, these search queries can be several pages of plaintext. Typically, these queries will include specific terms that need to be matched within documents, as well as terms to be excluded from documents. The queries often take a syntax similar to those of regular expressions; for example part of one query (out of possibly dozens of queries) could be:

\begin{center}
    \texttt{TITLE(cancer* or neoplasm* or neoplastic* or paraneoplas*)}
\end{center}

which would indicate a search for documents whose titles start with those particular terms. This query would be compiled with dozens of queries. Depending on the construction of the overall "meta" query, documents matching different combinations of the smaller queries would be returned. 

This proposed task centers around utilizing a Semantic IR system to replace the query construction and database search parts of the Systematic Review process, effectively replacing the "Identification" phase. The task calls for the development or application of an existing Semantic IR system that can take in a query written in (mostly or completely) natural language and fetches documents semantically relevant to the query, rather than having to manually construct complex search queries to retrieve such documents. Another added component in this process is multiple search query patterns that either refine or broaden previous ones (similar to a \texttt{AND} or \texttt{OR} operations, respectively). In contrast to this approach, the Semantic IR system adapted to the BioASQ Challenges described earlier in this work operates by accepting one query in natural language and returning results based on the single query. Such a system would need to be modified to accept multiple queries at once or a sequence of queries that would further refine the results in order to be suitable for this particular Living Systematic Review challenge.

As the order or ranking of the documents does not matter for the Systematic Review task, this challenge can be evaluated by metrics such as Precision, Recall, and F1 scores.

\subsubsection{Document Screening Task}
As described earlier, in the "Screening" phase of the Systematic Review pipeline, members of the team will generally manually review documents' title and abstract and make a determination whether the document should be excluded or remain in the pipeline to be further analysed in the next phase. This determination is made with respect to whether the title and abstract indicate that the document will meet the original intent of the particular Systematic Review, expressed through the original search queries.

This proposed task centers around using a Semantic IR system, or a system similar to or built upon the Semantic IR system, to classify whether a document (consisting of the title and abstract) should be excluded or kept in the pipeline based off the query. This objective is similar to that of the Semantic IR system detailed earlier in this work. The system applied to the BioASQ Challenge ranks documents and sentences given a query; the intended system in this proposal will classify documents given the query.

As mentioned previously, a Living Systematic Review pipeline will be designed to run at regular intervals on the same topic. Consequently, this system should be able to learn from repeated iterations of the pipeline on the same clinical topic and improve its performance on subsequent Systematic Review iterations.

Because the documents in this phase will be evaluated only on their title and abstracts, the same information that is used by IR systems participating in the BioASQ Challenge, it is possible to envision going directly from a query to the completion of the Screening phase. In other words, a single IR system could replace both the "Identification" and "Screening" phases in a Systematic Review pipeline that are usually manually conducted.

Similar to the proposed Document Identification Task, this task can be evaluated using the Precision, Recall, and F1 metrics.

\subsubsection{Document Eligibility Task}
The final task that we will propose is that of the Document Eligibility Task. The Document Eligibility step is considerably more challenging than the previous step in the pipeline of Screening. This is because when manually performed, this step involves reading through the entire document and locating information that justifies elimination or the keeping of the document. Moreover, there can be many reasons for elimination that may be present in the document, such as the size, scope, or quality of the clinical trials detailed in the document. Consequently, the criteria for keeping a document in the pipeline or excluding it will need to be explicitly specified for this task.

Whereas the previous proposed tasks have centered around document-level searching (specifically the title + abstract of a document, which has been evaluated previously in this work in the BioASQ Challenge), this particular task will require more detailed examination of the document. Moreover, correctly completing this task will necessitate developing modules that can correctly interpret nuances of concepts like clinical trials, as described in the examples above. Consequently, we anticipate this proposed challenge will be the most difficult to achieve effective results with.

\chapter{CONCLUSION AND FUTURE DIRECTIONS}
In this Chapter we will discuss the conclusion and avenues for further improvement in the future. The first section will discuss the results of the IR system developed in this work in the context of its performance on the BioASQ Challenge and its practical use. The next section will consist of additional improvements to the system and its applications that can be carried out as part of future work.

\section{Conclusions}
As discussed in previous Chapters, the system performed well on the BioASQ Sentence Retrieval challenges compared to other systems. Moreover, the Sentence Retrieval performance was notably high compared to the performance on the Document Retrieval task, of which the Sentence Retrieval task was downstream. This indicates that continuing to improve the Document Retrieval performance will further boost the Sentence Retrieval performance in the BioASQ Competition. 

As noted earlier in the paper, the primary focus of the research was on developing an IR system centered around Sentence Retrieval. This is reflected in the formulation of the Document Retrieval task as partially a Sentence Retrieval task. However, significant effort was not placed into document-specific modules that could potentially be used to boost the performance in the Document Retrieval task, a fact which will be discussed in the following Section.

The Alternating Optimization training loop appeared to empirically yield significant performance improvements. Specifically, conducting Bayesian Optimization across the parameter search space, and alternatively focusing on maximizing Sentence and Document MAP scores, seemed to provide a framework for boosting the performance for both tasks.

The Medic App is a proof-of-concept interface that demonstrates the abilities and value of using the system for practical scenarios. It was found that evaluating the system's performance objectively on the BioASQ Challenge was a good proxy for its observed performance in the Medic App. Consequently, continuing to improve the system for the BioASQ Challenge will increase its utility in the context of the Medic App.

We propose several different challenges regarding the Living Systematic Review task which will have significant practical benefits in the clinical domain. The systems developed in this work can contribute towards solving the challenges enumerated. 

One interesting area of consideration is the application of the "Multi-Perspective" nature of the Semantic IR system described in this work to other tasks or domains. In particular, the three major components, or "perspectives", used in this system -- Sentence Relevance, Semantic Textual Similarity, and Semantic Information Availability -- are well-suited for the Information Retrieval task, both intuitively and empirically. However, for other NLP tasks, or perhaps even tasks involving a fusion of modalities like vision or audio, alternative perspectives could be considered in the form of other benchmark scales. There is an interesting scope for consideration of the use of other novel perspectives, both in the context of Biomedical Information Retrieval, as well as in other domains or tasks.

\section{Future Work}
The Sentence Retrieval Multi-Perspective approach was shown to provide empirically better results than using the simple Sentence Relevance-based BERT model approach. Future work may potentially involve incorporating additional "perspectives" into the model, as well as additional lexical and word-based features, into the overall Sentence Retrieval module.

The Document Ranking module was formulated partially as a Sentence Ranking task; consequently, no document-specific modules besides the BM25 scores were considered in the module. To this end, focusing on document-centric modules, perhaps analogous to the modules considered for the Sentence Ranking tasks, would provide greater support for the Document Ranking task and improve scores. To this extent, methods that fall under a neural network-based approach similar to the Feedforward Fusion method as well as a Weighted Sum-type approach can be considered, but on the scale of entire documents rather than individual sentences. Doing so may provide a more cohesive perspective on the relevance of documents and enable the system to make more informed ranking decisions on the document level.

The Semantic Information Availability module contributed to the overall Semantic IR system; however, there is potential for improvement of the SIA systems by obtaining additional, higher-quality data. To this end, coming up with better heuristics and systems for the purpose of automatically generating SIA data will remain an area of future focus. With improved methodologies for automatically generating data, SIA systems will ideally continue improving in their predictions. Given the similarities between the SIA task and the overarching objective of an Information Retrieval system, improving SIA performance may lead to gains in associated IR systems as well.

Utilizing Bayesian Optimization across the parameter space in the Alternating Optimization training technique proved to be far more effective than utilizing gridsearch. However, the current implementation offers no guarantee the best parameters have been reached, either mathematically or empirically. Moreover, the importance of decisions such as initial parameter values, whether to perform early-stopping after a particular degree of convergence, and so on have yet to be considered. A thorough empirical evaluation of these different parameters will likely yield better insight into how this Alternating Optimization training loop can provide better sets of parameters for the Document Ranking and Sentence Ranking tasks. Additionally, simply spending more computation time in considering a greater number of hyperparameter combinations per training loop may yield immediate improved results.

Efforts will need to be taken to utilize the Semantic IR system detailed in this work for the Living Systematic Review challenges iterated in this work. The current challenges provide many avenues for the system detailed in this work to be applied; however, considerations specific to this particular domain and challenge will need to be considered.

The first area of focus with respect to applying Semantic IR to the Living Systematic Review task will be determining how to formulate the task of carrying out a search to retrieve documents based on an overarching research goal. The search queries mentioned earlier in this document are generally very nuanced and specific, and are constructed through a collaboration of domain experts like medical professionals and clinical librarians. Consequently, coming up with a way to incorporate this high degree of domain-specific knowledge into an IR system is the first area of focus. 

Next, there is a very close resemblance between the BioASQ task the system described in this work was focused around and the "Screening" phase of the Living Systematic Review task described. Consequently, once an effective means of formulating queries is identified, applying this system to that phase of the LSR pipeline will be considered. The major difference between the BioASQ task and the the "Screening" phase is that the BioASQ Challenges were based on ranking documents and sentences, while the "Screening" step is a task of binary classification. Therefore, the system will need to be adapted in order to perform this task.

All in all, there is scope for improvement of the various components of the Semantic IR system described in this work, as well as the system as a whole. There are also multiple areas of additional focus in terms of the applications of these systems.
{\singlespace
\addcontentsline{toc}{part}{REFERENCES}
\bibliographystyle{asudis}
\bibliography{dis}}
\renewcommand{\chaptername}{APPENDIX}
\addtocontents{toc}{APPENDIX \par}
\appendix
\chapter{BIOASQ RESULTS}
\newpage
The following are results from the BioASQ 6B and BioASQ 7B Competition.

\section{BioASQ 6B Document Ranking Results}

\begin{table}[h]
\begin{tabular}{|c|c|c|c|
>{\columncolor[HTML]{FFFC9E}}c |c|}
\hline
{System} & {\begin{tabular}[c]{@{}c@{}}MPrec\\ Docs\end{tabular}} & {\begin{tabular}[c]{@{}c@{}}MRec\\ Docs\end{tabular}} & {\begin{tabular}[c]{@{}c@{}}F-Measure\\ Docs\end{tabular}} & {\begin{tabular}[c]{@{}c@{}}MAP\\ Docs\end{tabular}} & {\begin{tabular}[c]{@{}c@{}}GMAP\\ Docs\end{tabular}} \\ \hline
aueb-nlp-3 & 0.3015 & 0.6387 & 0.3438 & 0.2327 & 0.0686 \\ \hline
aueb-nlp-4 & 0.2965 & 0.6220 & 0.3365 & 0.2288 & 0.0574 \\ \hline
aueb-nlp-2 & 0.2945 & 0.6307 & 0.3364 & 0.2275 & 0.0777 \\ \hline
aueb-nlp-1 & 0.2965 & 0.6260 & 0.3372 & 0.2265 & 0.0639 \\ \hline
\underline{Semantic IR} & \underline{0.2650} & \underline{0.5755} & \underline{0.3018} & \underline{0.2122} & \underline{0.0367} \\ \hline
sdm/rerank & 0.2500 & 0.5648 & 0.2886 & 0.1840 & 0.0277 \\ \hline
testtext & 0.2390 & 0.4206 & 0.2618 & 0.1742 & 0.0099 \\ \hline
ustb\_prir1 & 0.2390 & 0.4206 & 0.2618 & 0.1742 & 0.0099 \\ \hline
ustb\_prir3 & 0.2380 & 0.4193 & 0.2605 & 0.1738 & 0.0115 \\ \hline
ustb\_prir4 & 0.2380 & 0.4193 & 0.2605 & 0.1738 & 0.0115 \\ \hline
\end{tabular}
\caption[BioASQ 6B Phase A Batch 1 Document Ranking Results.]{BioASQ 6B Phase A Batch 1 Document Ranking results.}
\label{tab:bioasq6-batch1-doc}
\end{table}

\begin{table}[h]
\begin{tabular}{|c|c|c|c|
>{\columncolor[HTML]{FFFC9E}}c |c|}
\hline
{System} & {\begin{tabular}[c]{@{}c@{}}MPrec\\ Docs\end{tabular}} & {\begin{tabular}[c]{@{}c@{}}MRec\\ Docs\end{tabular}} & {\begin{tabular}[c]{@{}c@{}}F-Measure\\ Docs\end{tabular}} & {\begin{tabular}[c]{@{}c@{}}MAP\\ Docs\end{tabular}} & {\begin{tabular}[c]{@{}c@{}}GMAP\\ Docs\end{tabular}} \\ \hline
ustb\_prir3 & 0.3121 & 0.6379 & 0.3396 & 0.2512 & 0.0639 \\ \hline
ustb\_prir4 & 0.3121 & 0.6379 & 0.3396 & 0.2512 & 0.0639 \\ \hline
aueb-nlp-4 & 0.3220 & 0.6431 & 0.3479 & 0.2500 & 0.0660 \\ \hline
aueb-nlp-2 & 0.3210 & 0.6420 & 0.3475 & 0.2470 & 0.0701 \\ \hline
testtext & 0.3201 & 0.6355 & 0.3464 & 0.2467 & 0.0634 \\ \hline
ustb\_prir1 & 0.3201 & 0.6355 & 0.3464 & 0.2467 & 0.0634 \\ \hline
ustb\_prir2 & 0.3221 & 0.6618 & 0.3519 & 0.2458 & 0.0795 \\ \hline
aueb-nlp-3 & 0.3160 & 0.6365 & 0.3423 & 0.2416 & 0.0646 \\ \hline
aueb-nlp-1 & 0.3060 & 0.6294 & 0.3332 & 0.2319 & 0.0560 \\ \hline
\underline{Semantic IR} & \underline{0.2684} & \underline{0.5697} & \underline{0.2952} & \underline{0.2085} & \underline{0.0324} \\ \hline
\end{tabular}
\caption[BioASQ 6B Phase A Batch 2 Document Ranking Results.]{BioASQ 6B Phase A Batch 2 Document Ranking results.}
\label{tab:bioasq6-batch2-doc}
\end{table}

\begin{table}[h]
\begin{tabular}{|c|c|c|c|
>{\columncolor[HTML]{FFFC9E}}c |c|}
\hline
{System} & {\begin{tabular}[c]{@{}c@{}}MPrec\\ Docs\end{tabular}} & {\begin{tabular}[c]{@{}c@{}}MRec\\ Docs\end{tabular}} & {\begin{tabular}[c]{@{}c@{}}F-Measure\\ Docs\end{tabular}} & {\begin{tabular}[c]{@{}c@{}}MAP\\ Docs\end{tabular}} & {\begin{tabular}[c]{@{}c@{}}GMAP\\ Docs\end{tabular}} \\ \hline
ustb\_prir2 & 0.3340 & 0.6340 & 0.3396 & 0.2622 & 0.1057 \\ \hline
ustb\_prir3 & 0.3657 & 0.6244 & 0.3680 & 0.2610 & 0.0996 \\ \hline
testtext & 0.3637 & 0.6217 & 0.3655 & 0.2597 & 0.0991 \\ \hline
aueb-nlp-4 & 0.3577 & 0.6035 & 0.3545 & 0.2556 & 0.0811 \\ \hline
aueb-nlp-2 & 0.3557 & 0.5931 & 0.3511 & 0.2549 & 0.0796 \\ \hline
ustb\_prir4 & 0.3160 & 0.6203 & 0.3251 & 0.2506 & 0.0965 \\ \hline
ustb\_prir1 & 0.3250 & 0.6140 & 0.3314 & 0.2500 & 0.0923 \\ \hline
aueb-nlp-3 & 0.3477 & 0.5975 & 0.3462 & 0.2477 & 0.0844 \\ \hline
aueb-nlp-1 & 0.3437 & 0.5940 & 0.3432 & 0.2406 & 0.0824 \\ \hline
\underline{Semantic IR} & \underline{0.3020} & \underline{0.5664} & \underline{0.3065} & \underline{0.2386} & \underline{0.0647} \\ \hline
\end{tabular}
\caption[BioASQ 6B Phase A Batch 3 Document Ranking Results.]{BioASQ 6B Phase A Batch 3 Document Ranking results.}
\label{tab:bioasq6-batch3-doc}
\end{table}

\begin{table}[h]
\begin{tabular}{|c|c|c|c|
>{\columncolor[HTML]{FFFC9E}}c |c|}
\hline
{System} & {\begin{tabular}[c]{@{}c@{}}MPrec\\ Docs\end{tabular}} & {\begin{tabular}[c]{@{}c@{}}MRec\\ Docs\end{tabular}} & {\begin{tabular}[c]{@{}c@{}}F-Measure\\ Docs\end{tabular}} & {\begin{tabular}[c]{@{}c@{}}MAP\\ Docs\end{tabular}} & {\begin{tabular}[c]{@{}c@{}}GMAP\\ Docs\end{tabular}} \\ \hline
aueb-nlp-4 & 0.2446 & 0.5877 & 0.2751 & 0.1843 & 0.0208 \\ \hline
aueb-nlp-2 & 0.2456 & 0.5886 & 0.2760 & 0.1822 & 0.0207 \\ \hline
aueb-nlp-3 & 0.2466 & 0.5930 & 0.2779 & 0.1804 & 0.0218 \\ \hline
aueb-nlp-1 & 0.2436 & 0.5968 & 0.2749 & 0.1772 & 0.0225 \\ \hline
ustb\_prir1 & 0.2330 & 0.6244 & 0.2701 & 0.1665 & 0.0261 \\ \hline
ustb\_prir3 & 0.2459 & 0.6328 & 0.2821 & 0.1654 & 0.0281 \\ \hline
testtext & 0.2449 & 0.6327 & 0.2815 & 0.1650 & 0.0281 \\ \hline
ustb\_prir2 & 0.2450 & 0.6378 & 0.2827 & 0.1649 & 0.0275 \\ \hline
ustb\_prir4 & 0.2290 & 0.6015 & 0.2640 & 0.1638 & 0.0205 \\ \hline
... &  ... & ... & ... & ... & ... \\ \hline
\underline{Semantic IR} & \underline{0.1980} & \underline{0.5031} & \underline{0.2262} & \underline{0.1482} & \underline{0.0103} \\ \hline

\end{tabular}
\caption[BioASQ 6B Phase A Batch 4 Document Ranking Results.]{BioASQ 6B Phase A Batch 4 Document Ranking results.}
\label{tab:bioasq6-batch4-doc}
\end{table}

\begin{table}[h]
\begin{tabular}{|c|c|c|c|
>{\columncolor[HTML]{FFFC9E}}c |c|}
\hline
{System} & {\begin{tabular}[c]{@{}c@{}}MPrec\\ Docs\end{tabular}} & {\begin{tabular}[c]{@{}c@{}}MRec\\ Docs\end{tabular}} & {\begin{tabular}[c]{@{}c@{}}F-Measure\\ Docs\end{tabular}} & {\begin{tabular}[c]{@{}c@{}}MAP\\ Docs\end{tabular}} & {\begin{tabular}[c]{@{}c@{}}GMAP\\ Docs\end{tabular}} \\ \hline
aueb-nlp-4 & 0.2265 & 0.4424 & 0.2487 & 0.1464 & 0.0122 \\ \hline
aueb-nlp-3 & 0.2175 & 0.4200 & 0.2373 & 0.1422 & 0.0100 \\ \hline
aueb-nlp-2 & 0.2215 & 0.4356 & 0.2435 & 0.1416 & 0.0128 \\ \hline
aueb-nlp-1 & 0.2195 & 0.4197 & 0.2400 & 0.1403 & 0.0095 \\ \hline
sdm/rerank & 0.1975 & 0.4002 & 0.2230 & 0.1328 & 0.0074 \\ \hline
ustb\_prir2 & 0.2125 & 0.4073 & 0.2375 & 0.1278 & 0.0093 \\ \hline
ustb\_prir1 & 0.2090 & 0.3994 & 0.2312 & 0.1268 & 0.0072 \\ \hline
aueb-nlp-5 & 0.2702 & 0.3591 & 0.2552 & 0.1250 & 0.0077 \\ \hline
testtext & 0.2095 & 0.4070 & 0.2330 & 0.1248 & 0.0089 \\ \hline
... &  ... & ... & ... & ... & ... \\ \hline
\underline{Semantic IR} & \underline{0.1860} & \underline{0.3604} & \underline{0.2076} & \underline{0.1214} & \underline{0.0062} \\ \hline

\end{tabular}
\caption[BioASQ 6B Phase A Batch 5 Document Ranking Results.]{BioASQ 6B Phase A Batch 5 Document Ranking results.}
\label{tab:bioasq6-batch5-doc}
\end{table}

\clearpage
\section{BioASQ 6B Snippet Retrieval Results}

\begin{table}[h]
\begin{tabular}{|c|c|c|c|
>{\columncolor[HTML]{FFFC9E}}c |c|}
\hline
{System} & {\begin{tabular}[c]{@{}c@{}}MPrec\\ Docs\end{tabular}} & {\begin{tabular}[c]{@{}c@{}}MRec\\ Docs\end{tabular}} & {\begin{tabular}[c]{@{}c@{}}F-Measure\\ Docs\end{tabular}} & {\begin{tabular}[c]{@{}c@{}}MAP\\ Docs\end{tabular}} & {\begin{tabular}[c]{@{}c@{}}GMAP\\ Docs\end{tabular}} \\ \hline
\underline{Semantic IR} & \underline{0.2822} & \underline{0.3547} & \underline{0.2691} & \underline{0.2957} & \underline{0.0326} \\ \hline
aueb-nlp-3 & 0.2224 & 0.2593 & 0.2004 & 0.1684 & 0.0118 \\ \hline
aueb-nlp-1 & 0.2156 & 0.2496 & 0.1938 & 0.1675 & 0.0108 \\ \hline
aueb-nlp-4 & 0.2127 & 0.2501 & 0.1924 & 0.1659 & 0.0111 \\ \hline
aueb-nlp-2 & 0.2168 & 0.2599 & 0.1980 & 0.1642 & 0.0123 \\ \hline
ustb\_prir4 & 0.1634 & 0.1619 & 0.1364 & 0.1209 & 0.0015 \\ \hline
ustb\_prir2 & 0.1572 & 0.1616 & 0.1319 & 0.1205 & 0.0019 \\ \hline
ustb\_prir3 & 0.1691 & 0.1680 & 0.1393 & 0.1193 & 0.0018 \\ \hline
testtext & 0.1664 & 0.1723 & 0.1409 & 0.1151 & 0.0022 \\ \hline
ustb\_prir1 & 0.1664 & 0.1723 & 0.1409 & 0.1151 & 0.0022 \\ \hline
\end{tabular}
\caption[BioASQ 6B Phase A Batch 1 Snippet Ranking Results.]{BioASQ 6B Phase A Batch 1 Snippet Ranking results.}
\label{tab:bioasq6-batch1-sent}
\end{table}

\begin{table}[h]
\begin{tabular}{|c|c|c|c|
>{\columncolor[HTML]{FFFC9E}}c |c|}
\hline
{System} & {\begin{tabular}[c]{@{}c@{}}MPrec\\ Docs\end{tabular}} & {\begin{tabular}[c]{@{}c@{}}MRec\\ Docs\end{tabular}} & {\begin{tabular}[c]{@{}c@{}}F-Measure\\ Docs\end{tabular}} & {\begin{tabular}[c]{@{}c@{}}MAP\\ Docs\end{tabular}} & {\begin{tabular}[c]{@{}c@{}}GMAP\\ Docs\end{tabular}} \\ \hline
\underline{Semantic IR} & \underline{0.3476} & \underline{0.2946} & \underline{0.2399} & \underline{0.3335} & \underline{0.0230} \\ \hline
aueb-nlp-5 & 0.3852 & 0.2976 & 0.2653 & 0.3187 & 0.0352 \\ \hline
MindLab QA System ++ & 0.2878 & 0.2307 & 0.1985 & 0.2736 & 0.0065 \\ \hline
MindLab QA System & 0.2888 & 0.2298 & 0.1986 & 0.2695 & 0.0071 \\ \hline
MindLab QA Reloaded & 0.2888 & 0.2298 & 0.1986 & 0.2695 & 0.0071 \\ \hline
aueb-nlp-4 & 0.2873 & 0.2146 & 0.1850 & 0.2337 & 0.0231 \\ \hline
aueb-nlp-3 & 0.2746 & 0.2041 & 0.1749 & 0.2272 & 0.0210 \\ \hline
aueb-nlp-2 & 0.2768 & 0.2133 & 0.1826 & 0.2256 & 0.0236 \\ \hline
aueb-nlp-1 & 0.2716 & 0.2055 & 0.1749 & 0.2226 & 0.0202 \\ \hline
ustb\_prir4 & 0.2179 & 0.6188 & 0.2566 & 0.1731 & 0.0205 \\ \hline
\end{tabular}
\caption[BioASQ 6B Phase A Batch 2 Snippet Ranking Results.]{BioASQ 6B Phase A Batch 2 Snippet Ranking results.}
\label{tab:bioasq6-batch2-sent}
\end{table}

\begin{table}[h]
\begin{tabular}{|c|c|c|c|
>{\columncolor[HTML]{FFFC9E}}c |c|}
\hline
{System} & {\begin{tabular}[c]{@{}c@{}}MPrec\\ Docs\end{tabular}} & {\begin{tabular}[c]{@{}c@{}}MRec\\ Docs\end{tabular}} & {\begin{tabular}[c]{@{}c@{}}F-Measure\\ Docs\end{tabular}} & {\begin{tabular}[c]{@{}c@{}}MAP\\ Docs\end{tabular}} & {\begin{tabular}[c]{@{}c@{}}GMAP\\ Docs\end{tabular}} \\ \hline
\underline{Semantic IR} & \underline{0.3805} & \underline{0.3643} & \underline{0.3357} & \underline{0.3731} & \underline{0.0787} \\ \hline
aueb-nlp-5 & 0.3807 & 0.3655 & 0.3452 & 0.3320 & 0.0536 \\ \hline
aueb-nlp-4 & 0.2600 & 0.2607 & 0.2298 & 0.2306 & 0.0359 \\ \hline
aueb-nlp-1 & 0.2588 & 0.2612 & 0.2304 & 0.2278 & 0.0415 \\ \hline
aueb-nlp-3 & 0.2515 & 0.2525 & 0.2214 & 0.2221 & 0.0337 \\ \hline
MindLab QA Reloaded & 0.2374 & 0.2832 & 0.2377 & 0.2217 & 0.0134 \\ \hline
aueb-nlp-2 & 0.2532 & 0.2486 & 0.2211 & 0.2197 & 0.0319 \\ \hline
MindLab QA System & 0.2491 & 0.3094 & 0.2504 & 0.2113 & 0.0218 \\ \hline
testtext & 0.2951 & 0.2600 & 0.2452 & 0.2021 & 0.0246 \\ \hline
ustb\_prir3 & 0.2951 & 0.2600 & 0.2452 & 0.2021 & 0.0246 \\ \hline
\end{tabular}
\caption[BioASQ 6B Phase A Batch 3 Snippet Ranking Results.]{BioASQ 6B Phase A Batch 3 Snippet Ranking results.}
\label{tab:bioasq6-batch3-sent}
\end{table}

\begin{table}[h]
\begin{tabular}{|c|c|c|c|
>{\columncolor[HTML]{FFFC9E}}c |c|}
\hline
{System} & {\begin{tabular}[c]{@{}c@{}}MPrec\\ Docs\end{tabular}} & {\begin{tabular}[c]{@{}c@{}}MRec\\ Docs\end{tabular}} & {\begin{tabular}[c]{@{}c@{}}F-Measure\\ Docs\end{tabular}} & {\begin{tabular}[c]{@{}c@{}}MAP\\ Docs\end{tabular}} & {\begin{tabular}[c]{@{}c@{}}GMAP\\ Docs\end{tabular}} \\ \hline
\underline{Semantic IR} & \underline{0.2300} & \underline{0.3351} & \underline{0.2272} & \underline{0.2177} & \underline{0.0075} \\ \hline
aueb-nlp-5 & 0.2403 & 0.3229 & 0.2368 & 0.2138 & 0.0061 \\ \hline
aueb-nlp-2 & 0.1822 & 0.2566 & 0.1742 & 0.1555 & 0.0038 \\ \hline
aueb-nlp-3 & 0.1836 & 0.2513 & 0.1736 & 0.1480 & 0.0035 \\ \hline
aueb-nlp-4 & 0.1761 & 0.2367 & 0.1646 & 0.1479 & 0.0033 \\ \hline
aueb-nlp-1 & 0.1798 & 0.2377 & 0.1656 & 0.1444 & 0.0035 \\ \hline
MindLab QA System ++ & 0.1480 & 0.2342 & 0.1579 & 0.1413 & 0.0015 \\ \hline
ustb\_prir1 & 0.1752 & 0.3028 & 0.1753 & 0.1216 & 0.0077 \\ \hline
testtext & 0.1748 & 0.3027 & 0.1738 & 0.1213 & 0.0070 \\ \hline
ustb\_prir3 & 0.1749 & 0.3027 & 0.1739 & 0.1213 & 0.0070 \\ \hline
\end{tabular}
\caption[BioASQ 6B Phase A Batch 4 Snippet Ranking Results.]{BioASQ 6B Phase A Batch 4 Snippet Ranking results.}
\label{tab:bioasq6-batch4-sent}
\end{table}

\begin{table}[h]
\begin{tabular}{|c|c|c|c|
>{\columncolor[HTML]{FFFC9E}}c |c|}
\hline
{System} & {\begin{tabular}[c]{@{}c@{}}MPrec\\ Docs\end{tabular}} & {\begin{tabular}[c]{@{}c@{}}MRec\\ Docs\end{tabular}} & {\begin{tabular}[c]{@{}c@{}}F-Measure\\ Docs\end{tabular}} & {\begin{tabular}[c]{@{}c@{}}MAP\\ Docs\end{tabular}} & {\begin{tabular}[c]{@{}c@{}}GMAP\\ Docs\end{tabular}} \\ \hline
\underline{Semantic IR} & \underline{0.1450} & \underline{0.1902} & \underline{0.1325} & \underline{0.1219} & \underline{0.0018} \\ \hline
aueb-nlp-5 & 0.1360 & 0.1669 & 0.1248 & 0.1147 & 0.0020 \\ \hline
aueb-nlp-4 & 0.1409 & 0.1849 & 0.1285 & 0.1007 & 0.0024 \\ \hline
MindLab QA System & 0.0923 & 0.1444 & 0.0926 & 0.1006 & 0.0005 \\ \hline
MindLab QA Reloaded & 0.0908 & 0.1411 & 0.0911 & 0.1000 & 0.0005 \\ \hline
aueb-nlp-2 & 0.1364 & 0.1781 & 0.1238 & 0.0967 & 0.0028 \\ \hline
aueb-nlp-3 & 0.1303 & 0.1706 & 0.1175 & 0.0929 & 0.0021 \\ \hline
aueb-nlp-1 & 0.1362 & 0.1755 & 0.1224 & 0.0927 & 0.0024 \\ \hline
MindLab Red Lions++ & 0.0876 & 0.1446 & 0.0910 & 0.0874 & 0.0005 \\ \hline
ustb\_prir1 & 0.1264 & 0.1424 & 0.1100 & 0.0865 & 0.0010 \\ \hline
\end{tabular}
\caption[BioASQ 6B Phase A Batch 5 Snippet Ranking Results.]{BioASQ 6B Phase A Batch 5 Snippet Ranking results.}
\label{tab:bioasq6-batch5-sent}
\end{table}

\clearpage
\section{BioASQ 7B Document Ranking Results}

\begin{table}[h]
\begin{tabular}{|c|c|c|c|
>{\columncolor[HTML]{FFFC9E}}c |c|}
\hline
{System} & {\begin{tabular}[c]{@{}c@{}}MPrec\\ Docs\end{tabular}} & {\begin{tabular}[c]{@{}c@{}}MRec\\ Docs\end{tabular}} & {\begin{tabular}[c]{@{}c@{}}F-Measure\\ Docs\end{tabular}} & {\begin{tabular}[c]{@{}c@{}}MAP\\ Docs\end{tabular}} & {\begin{tabular}[c]{@{}c@{}}GMAP\\ Docs\end{tabular}} \\ \hline
\underline{Semantic IR} & \underline{0.1942} & \underline{0.5320} & \underline{0.2257} & \underline{0.1619} & \underline{0.0097} \\ \hline
Deep ML methods...  & 0.1930 & 0.5339 & 0.2230 & 0.1569 & 0.0089 \\ \hline
lh\_sys5 & 0.1950 & 0.5420 & 0.2288 & 0.1434 & 0.0084 \\ \hline
lh\_sys4 & 0.1910 & 0.5281 & 0.2237 & 0.1425 & 0.0075 \\ \hline
lh\_sys1 & 0.1910 & 0.5281 & 0.2237 & 0.1401 & 0.0078 \\ \hline
Ir\_sys2 & 0.1910 & 0.5443 & 0.2248 & 0.1388 & 0.0082 \\ \hline
lh\_sys2 & 0.1720 & 0.4534 & 0.1971 & 0.1300 & 0.0039 \\ \hline
Ir\_sys1 & 0.1770 & 0.5096 & 0.2095 & 0.1231 & 0.0067 \\ \hline
lh\_sys3 & 0.1770 & 0.4843 & 0.2053 & 0.1224 & 0.0048 \\ \hline
Ir\_sys4 & 0.1560 & 0.3949 & 0.1752 & 0.0926 & 0.0021 \\ \hline
\end{tabular}
\caption[BioASQ 7B Phase A Batch 1 Document Ranking Results.]{BioASQ 7B Phase A Batch 1 Document Ranking results.}
\label{tab:bioasq7-batch1-doc}
\end{table}

\begin{table}[h]
\begin{tabular}{|c|c|c|c|
>{\columncolor[HTML]{FFFC9E}}c |c|}
\hline
{System} & {\begin{tabular}[c]{@{}c@{}}MPrec\\ Docs\end{tabular}} & {\begin{tabular}[c]{@{}c@{}}MRec\\ Docs\end{tabular}} & {\begin{tabular}[c]{@{}c@{}}F-Measure\\ Docs\end{tabular}} & {\begin{tabular}[c]{@{}c@{}}MAP\\ Docs\end{tabular}} & {\begin{tabular}[c]{@{}c@{}}GMAP\\ Docs\end{tabular}} \\ \hline
aueb-nlp-4 & 0.2570 & 0.6342 & 0.2888 & 0.2181 & 0.0209 \\ \hline
aueb-nlp-2 & 0.2480 & 0.6159 & 0.2791 & 0.2056 & 0.0203 \\ \hline
aueb-nlp-5 & 0.4368 & 0.5861 & 0.4307 & 0.2032 & 0.0145 \\ \hline
aueb-nlp-1 & 0.2440 & 0.6292 & 0.2770 & 0.2009 & 0.0219 \\ \hline
aueb-nlp-3 & 0.2470 & 0.6335 & 0.2805 & 0.2007 & 0.0221 \\ \hline
\underline{Semantic IR} & \underline{0.2289} & \underline{0.5562} & \underline{0.2551} & \underline{0.1858} & \underline{0.0126} \\ \hline
lalala & 0.2260 & 0.5981 & 0.2610 & 0.1796 & 0.0132 \\ \hline
lh\_sys4 & 0.2290 & 0.5927 & 0.2613 & 0.1784 & 0.0118 \\ \hline
lh\_sys1 & 0.2260 & 0.5981 & 0.2610 & 0.1744 & 0.0127 \\ \hline
lh\_sys3 & 0.2260 & 0.5981 & 0.2610 & 0.1717 & 0.0123 \\ \hline
\end{tabular}
\caption[BioASQ 7B Phase A Batch 2 Document Ranking Results.]{BioASQ 7B Phase A Batch 2 Document Ranking results.}
\label{tab:bioasq7-batch2-doc}
\end{table}

\begin{table}[h]
\begin{tabular}{|c|c|c|c|
>{\columncolor[HTML]{FFFC9E}}c |c|}
\hline
{System} & {\begin{tabular}[c]{@{}c@{}}MPrec\\ Docs\end{tabular}} & {\begin{tabular}[c]{@{}c@{}}MRec\\ Docs\end{tabular}} & {\begin{tabular}[c]{@{}c@{}}F-Measure\\ Docs\end{tabular}} & {\begin{tabular}[c]{@{}c@{}}MAP\\ Docs\end{tabular}} & {\begin{tabular}[c]{@{}c@{}}GMAP\\ Docs\end{tabular}} \\ \hline
aueb-nlp-2 & 0.3610 & 0.6842 & 0.3952 & 0.2898 & 0.0841 \\ \hline
aueb-nlp-4 & 0.3460 & 0.6797 & 0.3820 & 0.2839 & 0.0862 \\ \hline
aueb-nlp-1 & 0.3420 & 0.6452 & 0.3730 & 0.2761 & 0.0663 \\ \hline
aueb-nlp-5 & 0.5476 & 0.6465 & 0.5289 & 0.2679 & 0.0740 \\ \hline
aueb-nlp-3 & 0.3310 & 0.6472 & 0.3658 & 0.2566 & 0.0657 \\ \hline
\underline{Semantic IR} & \underline{0.2901} & \underline{0.5782} & \underline{0.3230} & \underline{0.2443} & \underline{0.0346} \\ \hline
MindLab QA System & 0.2870 & 0.5778 & 0.3159 & 0.2392 & 0.0348 \\ \hline
MindLab Red Lions++ & 0.2870 & 0.5778 & 0.3159 & 0.2392 & 0.0348 \\ \hline
MindLab QA Reloaded & 0.2840 & 0.5709 & 0.3119 & 0.2323 & 0.0332 \\ \hline
Deep ML methods for & 0.2770 & 0.5658 & 0.3054 & 0.2272 & 0.0359 \\ \hline
\end{tabular}
\caption[BioASQ 7B Phase A Batch 3 Document Ranking Results.]{BioASQ 7B Phase A Batch 3 Document Ranking results.}
\label{tab:bioasq7-batch3-doc}
\end{table}

\begin{table}[h]
\begin{tabular}{|c|c|c|c|
>{\columncolor[HTML]{FFFC9E}}c |c|}
\hline
{System} & {\begin{tabular}[c]{@{}c@{}}MPrec\\ Docs\end{tabular}} & {\begin{tabular}[c]{@{}c@{}}MRec\\ Docs\end{tabular}} & {\begin{tabular}[c]{@{}c@{}}F-Measure\\ Docs\end{tabular}} & {\begin{tabular}[c]{@{}c@{}}MAP\\ Docs\end{tabular}} & {\begin{tabular}[c]{@{}c@{}}GMAP\\ Docs\end{tabular}} \\ \hline
aueb-nlp-1 & 0.2541 & 0.6668 & 0.2998 & 0.2102 & 0.0316 \\ \hline
aueb-nlp-2 & 0.2531 & 0.6523 & 0.2992 & 0.2092 & 0.0279 \\ \hline
aueb-nlp-4 & 0.2481 & 0.6445 & 0.2948 & 0.2080 & 0.0268 \\ \hline
aueb-nlp-5 & 0.4537 & 0.6416 & 0.4580 & 0.1968 & 0.0291 \\ \hline
aueb-nlp-3 & 0.2401 & 0.6451 & 0.2857 & 0.1962 & 0.0282 \\ \hline
lh\_sys4 & 0.2230 & 0.6121 & 0.2695 & 0.1752 & 0.0186 \\ \hline
\underline{Semantic IR} & \underline{0.2170} & \underline{0.5867} & \underline{0.2592} & \underline{0.1777} & \underline{0.0176} \\ \hline
MindLab QA Reloaded & 0.2080 & 0.5664 & 0.2463 & 0.1724 & 0.0121 \\ \hline
MindLab QA System ++ & 0.2080 & 0.5664 & 0.2463 & 0.1724 & 0.0121 \\ \hline
MindLab QA System & 0.2080 & 0.5664 & 0.2463 & 0.1724 & 0.0121 \\ \hline
\end{tabular}
\caption[BioASQ 7B Phase A Batch 4 Document Ranking Results.]{BioASQ 7B Phase A Batch 4 Document Ranking results.}
\label{tab:bioasq7-batch4-doc}
\end{table}

\begin{table}[h]
\begin{tabular}{|c|c|c|c|
>{\columncolor[HTML]{FFFC9E}}c |c|}
\hline
{System} & {\begin{tabular}[c]{@{}c@{}}MPrec\\ Docs\end{tabular}} & {\begin{tabular}[c]{@{}c@{}}MRec\\ Docs\end{tabular}} & {\begin{tabular}[c]{@{}c@{}}F-Measure\\ Docs\end{tabular}} & {\begin{tabular}[c]{@{}c@{}}MAP\\ Docs\end{tabular}} & {\begin{tabular}[c]{@{}c@{}}GMAP\\ Docs\end{tabular}} \\ \hline
aueb-nlp-2 & 0.1570 & 0.4221 & 0.1802 & 0.1218 & 0.0036 \\ \hline
aueb-nlp-4 & 0.1580 & 0.4632 & 0.1905 & 0.1080 & 0.0052 \\ \hline
aueb-nlp-1 & 0.1510 & 0.4346 & 0.1799 & 0.1049 & 0.0037 \\ \hline
aueb-nlp-5 & 0.2708 & 0.4196 & 0.2712 & 0.1004 & 0.0038 \\ \hline
aueb-nlp-3 & 0.1440 & 0.4406 & 0.1757 & 0.0968 & 0.0042 \\ \hline
lh\_sys5 & 0.1350 & 0.4432 & 0.1653 & 0.0884 & 0.0033 \\ \hline
lh\_sys4 & 0.1310 & 0.4311 & 0.1612 & 0.0850 & 0.0028 \\ \hline
lh\_sys1 & 0.1280 & 0.4235 & 0.1570 & 0.0829 & 0.0028 \\ \hline
Deep ML methods... & 0.1240 & 0.4004 & 0.1497 & 0.0823 & 0.0026 \\ \hline
.. & .. & .. & .. & .. & ..\\ \hline
\underline{Semantic IR} & \underline{0.1180} & \underline{0.3659} & \underline{0.1440} & \underline{0.0806} & \underline{0.0019} \\ \hline
\end{tabular}
\caption[BioASQ 7B Phase A Batch 5 Document Ranking Results.]{BioASQ 7B Phase A Batch 5 Document Ranking results.}
\label{tab:bioasq7-batch5-doc}
\end{table}

\clearpage
\section{BioASQ 7B Sentence Ranking Results}

\begin{table}[h]
\begin{tabular}{|c|c|c|c|
>{\columncolor[HTML]{FFFC9E}}c |c|}
\hline
{System} & {\begin{tabular}[c]{@{}c@{}}MPrec\\ Docs\end{tabular}} & {\begin{tabular}[c]{@{}c@{}}MRec\\ Docs\end{tabular}} & {\begin{tabular}[c]{@{}c@{}}F-Measure\\ Docs\end{tabular}} & {\begin{tabular}[c]{@{}c@{}}MAP\\ Docs\end{tabular}} & {\begin{tabular}[c]{@{}c@{}}GMAP\\ Docs\end{tabular}} \\ \hline
\underline{Semantic IR} & \underline{0.2134} & \underline{0.3702} & \underline{0.2426} & \underline{0.2112} & \underline{0.0060} \\ \hline
Deep ML methods... & 0.1529 & 0.2933 & 0.1773 & 0.1411 & 0.0029 \\ \hline
\end{tabular}
\caption[BioASQ 7B Phase A Batch 1 Snippet Ranking Results.]{BioASQ 7B Phase A Batch 1 Snippet Ranking results.}
\emph{Note: only one other participant for this batch.}
\label{tab:bioasq7-batch1-sent}
\end{table}

\begin{table}[h]
\begin{tabular}{|c|c|c|c|
>{\columncolor[HTML]{FFFC9E}}c |c|}
\hline
{System} & {\begin{tabular}[c]{@{}c@{}}MPrec\\ Docs\end{tabular}} & {\begin{tabular}[c]{@{}c@{}}MRec\\ Docs\end{tabular}} & {\begin{tabular}[c]{@{}c@{}}F-Measure\\ Docs\end{tabular}} & {\begin{tabular}[c]{@{}c@{}}MAP\\ Docs\end{tabular}} & {\begin{tabular}[c]{@{}c@{}}GMAP\\ Docs\end{tabular}} \\ \hline
aueb-nlp-5 & 0.2746 & 0.3414 & 0.2623 & 0.2650 & 0.0099 \\ \hline
aueb-nlp-1 & 0.2536 & 0.3785 & 0.2535 & 0.2609 & 0.0100 \\ \hline
aueb-nlp-2 & 0.2448 & 0.3763 & 0.2454 & 0.2449 & 0.0110 \\ \hline
\underline{Semantic IR} & \underline{0.2394} & \underline{0.3117} & \underline{0.2228} & \underline{0.2435} & \underline{0.0067} \\ \hline
aueb-nlp-4 & 0.1955 & 0.2652 & 0.1805 & 0.1820 & 0.0066 \\ \hline
Deep ML methods for & 0.1743 & 0.2354 & 0.1752 & 0.1750 & 0.0017 \\ \hline
aueb-nlp-3 & 0.1892 & 0.2723 & 0.1777 & 0.1663 & 0.0062 \\ \hline
lh\_sys3 & 0.1269 & 0.1708 & 0.1196 & 0.0935 & 0.0014 \\ \hline
lh\_sys5 & 0.1276 & 0.1712 & 0.1201 & 0.0933 & 0.0013 \\ \hline
lh\_sys1 & 0.1269 & 0.1708 & 0.1196 & 0.0894 & 0.0014 \\ \hline
\end{tabular}
\caption[BioASQ 7B Phase A Batch 2 Snippet Ranking Results.]{BioASQ 7B Phase A Batch 2 Snippet Ranking results.}
\label{tab:bioasq7-batch2-sent}
\end{table}

\begin{table}[h]
\begin{tabular}{|c|c|c|c|
>{\columncolor[HTML]{FFFC9E}}c |c|}
\hline
{System} & {\begin{tabular}[c]{@{}c@{}}MPrec\\ Docs\end{tabular}} & {\begin{tabular}[c]{@{}c@{}}MRec\\ Docs\end{tabular}} & {\begin{tabular}[c]{@{}c@{}}F-Measure\\ Docs\end{tabular}} & {\begin{tabular}[c]{@{}c@{}}MAP\\ Docs\end{tabular}} & {\begin{tabular}[c]{@{}c@{}}GMAP\\ Docs\end{tabular}} \\ \hline
aueb-nlp-2 & 0.3933 & 0.4120 & 0.3522 & 0.3864 & 0.0777 \\ \hline
aueb-nlp-1 & 0.3636 & 0.3755 & 0.3219 & 0.3657 & 0.0454 \\ \hline
aueb-nlp-5 & 0.3678 & 0.3609 & 0.3277 & 0.3481 & 0.0533 \\ \hline
\underline{Semantic IR} & \underline{0.3323} & \underline{0.3333} & \underline{0.2879} & \underline{0.3470} & \underline{0.0315} \\ \hline
aueb-nlp-4 & 0.2853 & 0.2735 & 0.2454 & 0.2609 & 0.0335 \\ \hline
aueb-nlp-3 & 0.2715 & 0.2661 & 0.2373 & 0.2349 & 0.0266 \\ \hline
MindLab QA Reloaded & 0.2202 & 0.2541 & 0.2046 & 0.2330 & 0.0061 \\ \hline
Deep ML methods for & 0.2049 & 0.2596 & 0.2052 & 0.1955 & 0.0083 \\ \hline
MindLab Red Lions++ & 0.2072 & 0.2178 & 0.1804 & 0.1820 & 0.0135 \\ \hline
MindLab QA System & 0.1874 & 0.1980 & 0.1675 & 0.1690 & 0.0057 \\ \hline
\end{tabular}
\caption[BioASQ 7B Phase A Batch 3 Snippet Ranking Results.]{BioASQ 7B Phase A Batch 3 Snippet Ranking results.}
\label{tab:bioasq7-batch3-sent}
\end{table}

\begin{table}[h]
\begin{tabular}{|c|c|c|c|
>{\columncolor[HTML]{FFFC9E}}c |c|}
\hline
{System} & {\begin{tabular}[c]{@{}c@{}}MPrec\\ Docs\end{tabular}} & {\begin{tabular}[c]{@{}c@{}}MRec\\ Docs\end{tabular}} & {\begin{tabular}[c]{@{}c@{}}F-Measure\\ Docs\end{tabular}} & {\begin{tabular}[c]{@{}c@{}}MAP\\ Docs\end{tabular}} & {\begin{tabular}[c]{@{}c@{}}GMAP\\ Docs\end{tabular}} \\ \hline
aueb-nlp-2 & 0.3254 & 0.4308 & 0.3048 & 0.3409 & 0.0344 \\ \hline
aueb-nlp-1 & 0.3209 & 0.4321 & 0.3018 & 0.3249 & 0.0281 \\ \hline
\underline{Semantic IR} & \underline{0.2959} & \underline{0.3414} & \underline{0.2568} & \underline{0.3030} & \underline{0.0151} \\ \hline
aueb-nlp-5 & 0.3256 & 0.4403 & 0.3010 & 0.2976 & 0.0379 \\ \hline
MindLab QA Reloaded & 0.2276 & 0.2857 & 0.2093 & 0.2214 & 0.0052 \\ \hline
aueb-nlp-3 & 0.2563 & 0.3581 & 0.2346 & 0.2213 & 0.0196 \\ \hline
aueb-nlp-4 & 0.2550 & 0.3325 & 0.2318 & 0.2173 & 0.0178 \\ \hline
MindLab Red Lions++ & 0.2168 & 0.2718 & 0.1982 & 0.2000 & 0.0067 \\ \hline
MindLab QA System ++ & 0.2112 & 0.2317 & 0.1819 & 0.1931 & 0.0058 \\ \hline
MindLab QA System & 0.1998 & 0.2669 & 0.1865 & 0.1892 & 0.0064 \\ \hline
\end{tabular}
\caption[BioASQ 7B Phase A Batch 4 Snippet Ranking Results.]{BioASQ 7B Phase A Batch 4 Snippet Ranking results.}
\label{tab:bioasq7-batch4-sent}
\end{table}

\begin{table}[h]
\begin{tabular}{|c|c|c|c|
>{\columncolor[HTML]{FFFC9E}}c |c|}
\hline
{System} & {\begin{tabular}[c]{@{}c@{}}MPrec\\ Docs\end{tabular}} & {\begin{tabular}[c]{@{}c@{}}MRec\\ Docs\end{tabular}} & {\begin{tabular}[c]{@{}c@{}}F-Measure\\ Docs\end{tabular}} & {\begin{tabular}[c]{@{}c@{}}MAP\\ Docs\end{tabular}} & {\begin{tabular}[c]{@{}c@{}}GMAP\\ Docs\end{tabular}} \\ \hline
aueb-nlp-2 & 0.1459 & 0.3019 & 0.1575 & 0.1383 & 0.0018 \\ \hline
aueb-nlp-5 & 0.1391 & 0.2830 & 0.1520 & 0.1194 & 0.0016 \\ \hline
aueb-nlp-1 & 0.1168 & 0.2681 & 0.1312 & 0.1146 & 0.0010 \\ \hline
\underline{Semantic IR} & \underline{0.1103} & \underline{0.2500} & \underline{0.1292} & \underline{0.1122} & \underline{0.0013} \\ \hline
aueb-nlp-4 & 0.1118 & 0.2600 & 0.1222 & 0.0948 & 0.0013 \\ \hline
aueb-nlp-3 & 0.1033 & 0.2361 & 0.1126 & 0.0859 & 0.0010 \\ \hline
MindLab QA System ++ & 0.0720 & 0.1923 & 0.0874 & 0.0781 & 0.0004 \\ \hline
MindLab QA Reloaded & 0.0720 & 0.1923 & 0.0874 & 0.0781 & 0.0004 \\ \hline
MindLab QA System & 0.0737 & 0.2129 & 0.0885 & 0.0670 & 0.0005 \\ \hline
MindLab Red Lions++ & 0.0701 & 0.1970 & 0.0854 & 0.0578 & 0.0005 \\ \hline
\end{tabular}
\caption[BioASQ 7B Phase A Batch 5 Snippet Ranking Results.]{BioASQ 7B Phase A Batch 5 Snippet Ranking results.}
\label{tab:bioasq7-batch5-sent}
\end{table}

\clearpage

\chapter{LINKS TO CODE REPOSITORY}
\newpage
The code for this work is available at https://github.com/samrawal/bio-semantic-ir.

\include{vita}
\end{document}